\documentclass[%
 reprint,
superscriptaddress,
 amsmath,amssymb,
 aps,
]{revtex4-2}

\usepackage{graphicx}
\usepackage{dcolumn}
\usepackage{bm}

\begin{document}

\title{\textit{Ab initio} Simulation of Laser-Induced Electronic and Vibrational Coherence}
\author{Jannis Krumland}
\email{jannis.krumland@physik.hu-berlin.de}
\affiliation{Physics Department and IRIS Adlershof, Humboldt-Universität zu Berlin, 12489 Berlin, Germany}
\author{Matheus Jacobs}
\affiliation{Physics Department and IRIS Adlershof, Humboldt-Universität zu Berlin, 12489 Berlin, Germany}
\author{Caterina Cocchi}
\email{caterina.cocchi@uni-oldenburg.de}
\affiliation{Physics Department and IRIS Adlershof, Humboldt-Universität zu Berlin, 12489 Berlin, Germany}
\affiliation{Institute of Physics, Carl von Ossietzky Universität Oldenburg, 26129 Oldenburg, Germany}

\date{\today}

\begin{abstract}
The atomistic resolution recently achieved by ultrafast spectroscopies demands corresponding theoretical advances. Real-time time-dependent density-functional theory (RT-TDDFT) with Ehrenfest dynamics offers an optimal trade-off between accuracy and computational costs to study electronic and vibrational dynamics of laser-excited materials in the sub-picosecond regime. However, this approach is unable to account for thermal effects or zero-point energies which are crucial in the physics involved. Herein, we adopt a quantum-semiclassical method based on RT-TDDFT+Ehrenfest to simulate laser-induced electronic and vibrational coherences in condensed matter. With the example of carbon-conjugated molecules, we show that ensemble-averaging with initial configurations from a nuclear quantum distribution remedies many shortcomings of single-trajectory RT-TDDFT+Ehrenfest, damping electronic coherence and introducing ultrafast non-adiabatic coupling between excited states. As the number of sampled configurations decreases with size and rigidity of the compounds, computational costs remain moderate for large systems for which mean-field approaches shine. The explicit inclusion of a time-dependent pulse in the simulations makes this method a prime advance for first-principles studies of coherent nonlinear spectroscopy as an independent counterpart to experimental results.
\end{abstract}

\maketitle

\newpage

\section{Introduction}

State-of-the-art nonlinear spectroscopy techniques finally enable   
tracking ultrafast coupled electronic-vibrational motion on its natural timescale, granting unprecedented insight into the quantum-mechanical effects ruling nanoscale physical phenomena~\cite{mukamel2000arpc, jonas2003arpc, cho2008cr, hamm+zanni2011,coll21jpcc}. These advances demand time-resolved first-principles simulations complementary to measurements~\cite{conti+20jacs}. Within the variety of available approaches to non-adiabatic \textit{ab initio} molecular dynamics~\cite{curc+mart2018cr}, mean-field techniques~\cite{tully1998faraday} are appealing thanks to their excellent numerical scalability. Among them, the Ehrenfest scheme -- the classical-nuclei limit of the time-dependent self-consistent field approximation~\cite{heller1976jcp_tdscf} -- is particularly attractive as it can be straightforwardly coupled to real-time time-dependent density functional theory (RT-TDDFT), whereby the quantum-mechanical electronic subsystem explicitly evolves in time. In the last decade, RT-TDDFT+Ehrenfest has successfully complemented experiments to rationalize charge-transfer dynamics in realistic complexes~\cite{rozzi2013natcom, falke+2014sci, rozzi+2017jpcm, zhang+2017as, jaco+2020advpx}. 
 
The direct time propagation of the electronic subsystem in RT-TDDFT enables the explicit inclusion in the simulation of a laser field elevating the system to a non-stationary excited state (ES), in close similarity with the experimental scenario~\cite{desi-lien17pccp}. All electronically coherent features arising from the system not being in an eigenstate of its Hamiltonian are naturally captured, including linear-response polarization, as well as quantum interferences between different ES.  
However, in this context, there are some problems associated with the complete lack of electron-nuclei correlation in the single-trajectory RT-TDDFT+Ehrenfest (STE) scheme. Laser-initiated electron dynamics remain overly coherent when the nuclei have always well-defined positions and momenta. Moreover, neglecting zero-point energy (ZPE) is hardly justifiable in organic systems, where it is much higher than thermal energy due to the low atomic masses.

Here, we analyze to what extent a multi-trajectory RT-TDDFT+Ehrenfest (MTE) approach with random initial configurations from quantum distributions can overcome the shortcomings of single-trajectory calculations. The formalism retrieves some of the electron-nuclear correlation that is lost in making the time-dependent self-consistent field approximation, the quantum-mechanical parent of single-trajectory Ehrenfest molecular dynamics~\cite{nancy+miller1987}.
We account for coherent electron-vibrational couplings in systems excited by a laser pulse of defined shape, polarization, intensity, and duration. 
Results obtained for prototypical carbon-conjugated molecules, namely benzene and coronene, are contrasted against corresponding single-trajectory simulations and ensemble-averaged calculations with fixed nuclei. This comparison highlights the role of nuclear motion, which redistributes the oscillator strength in optical spectra through non-adiabatic couplings between ES. These processes are particularly evident when considering nonlinear response. The population dynamics are mainly encoded in the second order; results of corresponding simulations resemble established methods for accessing non-adiabatic dynamics, but naturally feature also transient contributions related to coherences between ES. Finally, we consider the nuclear motion triggered by electronic excitations, demonstrating the ability of the proposed approach to simulate wavepacket motion, which follows an almost classical time evolution for fully symmetric modes, while becoming non-trivial for the less symmetric ones.

\section{Methodology}

MTE simulations are initialized by generating a set of nuclear coordinates and velocities. The coordinates ${\cal Q}_\alpha$ and the momenta ${\cal P}_\alpha$   in the normal-mode basis are randomly sampled from a Wigner distribution, which - for harmonic oscillators - reads~\cite{hillery+1984pr}
\begin{align}\label{eq.wigner}
\Gamma(\{{\cal Q}_\alpha\}, \{{\cal P}_\alpha\}) \propto \prod_\alpha\, f^{(p)}_\alpha({\cal P}_\alpha)f^{(q)}_\alpha({\cal Q}_\alpha),
\end{align}
where $f^{(p)}_\alpha$ and $f^{(q)}_\alpha$ are zero-centered Gaussians with standard deviations
\begin{subequations}\label{eq.stddev}
\begin{align}
\sigma_\alpha^{(p)} &= \left(\frac{\hbar\Omega_\alpha}{2\tanh(\beta\hbar\Omega_\alpha/2)}\right)^{1/2}\hspace{0.3cm}\text{and}\\
\sigma_\alpha^{(q)} &= \left(\frac{\hbar}{2\Omega_\alpha\tanh(\beta\hbar\Omega_\alpha/2)}\right)^{1/2},
\end{align}
\end{subequations}
respectively, where $\beta = 1/k_\mathrm{B}T$, with $T$ being the temperature, and $\Omega_\alpha$ is the frequency of the vibrational mode $\alpha$. 
Normal coordinates and momenta sampled from this distribution are transformed into Cartesian initial conditions,
\begin{subequations}
\begin{align}\label{eq.normalTrafo}
    R_{K\nu}(t=0) &= M_K^{-1/2}\sum_{\alpha}{\cal T}^{-1}_{\alpha,K\nu}{\cal Q}_\alpha \\ \frac{\text dR_{K\nu}}{\text dt}(t=0)  &= M_K^{-1/2}\sum_{\alpha}{\cal T}^{-1}_{\alpha,K\nu}{\cal P}_\alpha,
\end{align}
\end{subequations}
where $\textbf{R}_K$ and $M_K$ are position and mass of the $K$-th nucleus, respectively, and ${\cal T}_{\alpha,K\nu}$ is the matrix transforming mass-weighted Cartesian coordinates into normal ones. The normal frequencies $\Omega_\alpha$ and the transformation matrix ${\cal T}_{\alpha,K\nu}$ required for the generation of these starting configurations are obtained from density-functional perturbation theory. Subsequently, the nuclear subsystem evolves based on the classical forces acting on them~\cite{ullrich2012oxford}:
\begin{align}\label{eq.ehrenfest}
    &M_K\frac{\text d^2\mathbf{R}_K}{\text dt^2} =\nonumber\\
     &-\left.\nabla_{\textbf{R}_K}\left[V_{nn}(R)+\int\text d^3r\,\rho(\textbf{r},t)V_{en}(\textbf{r},R)\right]\right|_{R=R(t)},
\end{align}
where $R = \lbrace\textbf{R}_K\rbrace$ is the set of the positions of all nuclei, $V_{\mathrm{nn}}$ is the electrostatic repulsion between them, and $\rho$ is the electron density,
\begin{align}
\rho(\textbf{r},t) = \sum_n^{\text{occ}}|\psi_n(\textbf{r},t)|^2,
\end{align}
which is calculated from the occupied time-dependent Kohn-Sham orbitals, $\psi_n$.
The time evolution of these orbitals is performed, using RT-TDDFT~\cite{rung+1984prl}, starting from a ground-state density obtained from density functional theory~\cite{hohenbergKohn1964pr,kohnSham1965pr}. The electronic equation of motion is the time-dependent Kohn-Sham equation,
\begin{align}\label{eq.ks}
    i\hbar\frac{\partial}{\partial t}\psi(\textbf{r},t) = \hat{\cal H}_{\text{KS}}[\rho](\textbf{r},t)\psi(\textbf{r},t).
\end{align}
The Kohn-Sham Hamiltonian in Eq.~\eqref{eq.ks},
\begin{align}\label{eq.ks_ham}
    \hat{\cal H}_{\text{KS}}[\rho](\textbf{r},t) &= -\frac{\hbar^2}{2m_e}\nabla^2+ V_\mathrm{en}(\textbf{r},R(t))+V_{\text{ext}}(\textbf{r},t)\nonumber\\&+V_\text{H}[\rho(t)](\textbf{r},t)+V_\text{xc}[\rho](\textbf{r},t),
\end{align}
contains the kinetic energy, the electrostatic potential generated by the nuclei, $V_\mathrm{en}$, and the interaction with the external potential, $V_{\text{ext}}$, arising from the coupling to a Gaussian-enveloped, dynamical electric field in the dipole approximation,
\begin{align}\label{eq.electricField}
    V_{\text{ext}}(\textbf{r},t) &= e\textbf{r}\cdot\textbf{E}(t) \nonumber\\&= e\textbf{r}\cdot\hat{\mathbf{n}}E_0\exp(-(t-t_\mu)^2/2t_\sigma^2)\cos(\omega_pt),
\end{align}
which is characterized by the polarization direction $\hat{\mathbf{n}}$, the field strength $E_0$, the pulse center $t_\mu$, the width $t_\sigma$, and the carrier frequency $\omega_p$. Finally, the last two terms in Eq.~\eqref{eq.ks_ham}, carrying a functional dependence on $\rho$, describe interactions among electrons, 
including the Hartree potential, $V_\mathrm{H}[\rho(t)]$, and the exchange-correlation one, $V_\mathrm{xc}[\rho]$. Much of the complexity of the dynamical many-electron problem is contained in $V_\mathrm{xc}[\rho]$, the exact form of which is unknown and requires approximations. Two layers of approximation are usually made: (i) the neglect of memory, \textit{i.e.} the dependence on $\rho(t'<t)$, the electron density at earlier times~\cite{mait+2002prl}, and (ii) the approximation of the instantaneous $V_\mathrm{xc}[\rho(t)]$ by a ground-state density functional, inserting $\rho(t)$ instead of the ground-state density. The adiabatic approximation (i) is most accurate when the electronic system remains close to the ground state~\cite{lacombe2020fd, lacombe+maitra2021jpcl}, such that it is advisable to choose a small laser amplitude, $E_0$, generating only little excited-state population.

All simulations are carried out with version 9.2 of the \textsc{Octopus} code~\cite{octopus2015, octopus2020}. Wavefunctions are represented on a real-space grid generated by sampling with a spacing of 0.24~\AA\,the union of spheres of radius 5~\AA\,centered at each atomic site. Using the FIRE algorithm~\cite{fire}, geometries are optimized until forces are below 10$^{-3}$~eV/\AA~before determining the normal modes of vibration. 500 initial geometries and velocities are generated by sampling Eq.~\eqref{eq.wigner} for all modes (excluding rotations and translations). For the ensuing time evolution, we employ the approximated enforced time-reversal symmetry scheme~\cite{castro+2004jcp} with a time step of 2.7~as. The Perdew-Zunger variant~\cite{pz} of the adiabatic local-density approximation is used for the exchange-correlation potential, and nuclear potentials are described with Troullier-Martins pseudopotentials~\cite{trou-mart91prb}. The parameters for the electric field in Eq.~\eqref{eq.electricField} are chosen as $\hat{\textbf{n}} = (1,1,1)/\sqrt{3}$, $t_\mu$~=~8~fs, $t_\sigma$~=~2~fs, and $E_0$ corresponding to a peak intensity of about 3.5$\times$10$^{10}$~W/cm$^2$. The carrier frequency $\omega_p$ is set to 6.9~eV, 3.7~eV, and 3.9~eV for benzene, coronene, and N-substituted coronene, respectively. No thermal energy is added, \textit{i.e.}, $T=0$, though test runs reveal that the difference between $T=0$ and $T=300$K is rather small (Fig.~S2). 

\section{Results and Discussion}
\subsection{Electron Dynamics: Linear Regime}
\begin{figure*}
    \centering
    \includegraphics[width=.95\textwidth]{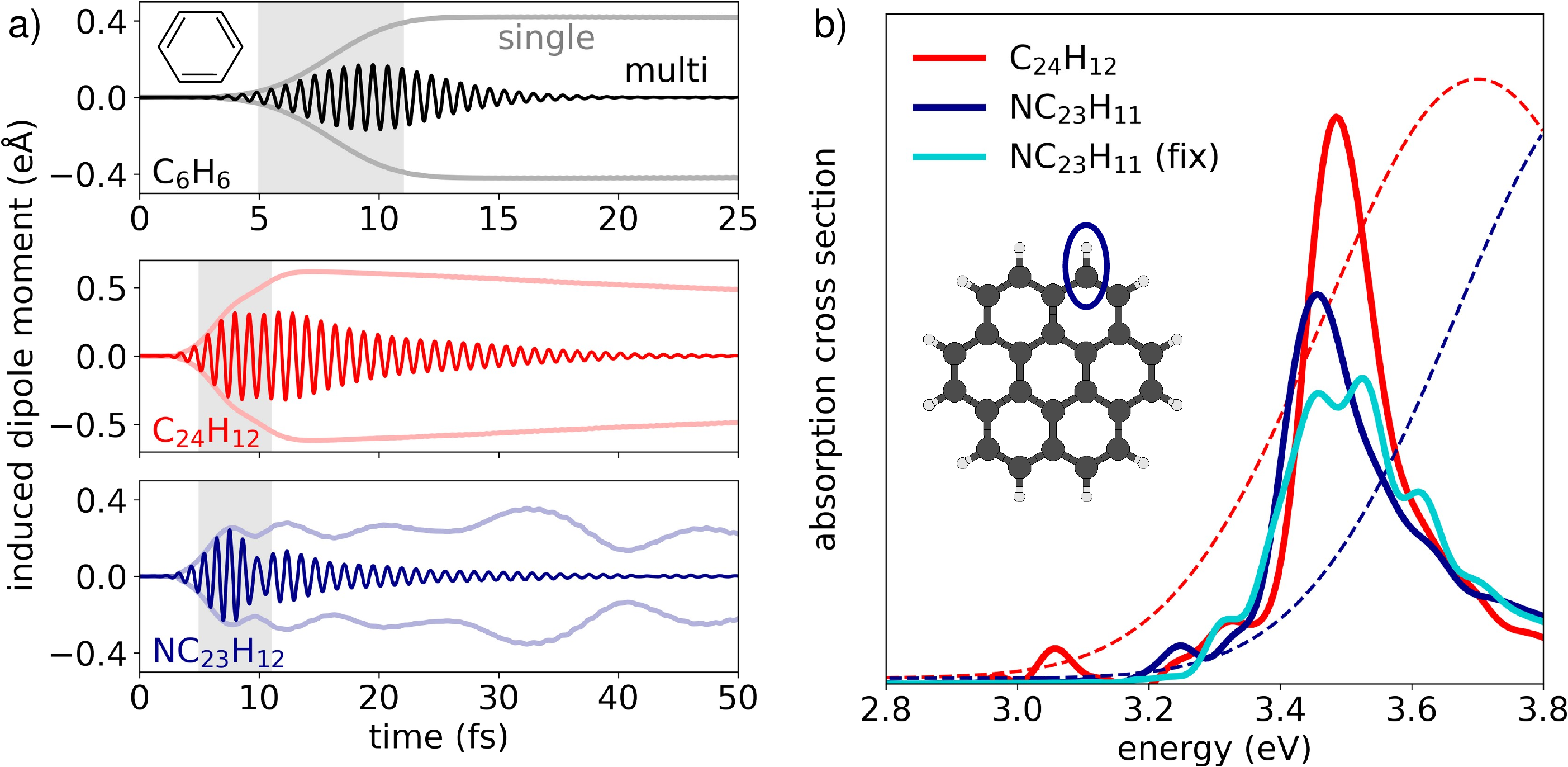}
    \caption{a) Dipole moment induced by a pulse (grey area) resonantly exciting benzene (top), coronene (middle), and N-substituted coronene (bottom). Faint curves represent the envelopes resulting from single-trajectory calculations starting from rest. b) Absorption spectrum of coronene (red curve) and its N-substituted counterpart (blue) triggered by pulses with spectra shown by dashed lines. The same ensemble is used for N-coronene with nuclei fixed in their initial configuration (turquoise). Inset: Ball-and-stick representation of coronene (C atoms in grey, H in white), with the CH group circled in blue replaced by N in the substituted counterpart.}
    \label{fig.1}
\end{figure*}

The central quantity to consider in the analysis of the electron dynamics is the  ensemble-averaged electronic dipole moment. This quantity creates a polarization in macroscopic samples, which, in turn, gives rise to an emitted electric field~\cite{krum+21jcp}; oscillations of the induced dipole moment are an indicator of (electronic) coherence. In benzene, where the linear absorption spectrum~\cite{kochOtto1972cpl} is reproduced remarkably well by the employed adiabatic local-density approximation~\cite{yabanaBertsch1999ijqp}, a laser pulse in resonance with the strong $1^1E_{1u}\leftarrow 1^1A_{1g}$ excitation at 6.9~eV causes merely short-lived dipolar oscillations [Fig.~\ref{fig.1}a), top]: A monoexponential fit to the decaying envelope yields a dephasing time $T_{01}$ = 2.5~fs. The corresponding first-order density matrix after the pulse, giving rise to the induced dipole moment, is
\begin{align}\label{eq.linear_single}
    \hat\rho^{(1)}(t) \sim \exp(i\omega_{01}t-t/T_{01})|0\rangle\langle 1|,
\end{align}
where $|0\rangle$ and $|1\rangle$ represent $1^1A_{1g}$ and $1^1E_{1u}$ states with energies $E_0$ and $E_1$, respectively, and $\omega_{01} = (E_1-E_0)/\hbar$. In contrast to the exponential decay, no damping is observed in the single-trajectory case (grey faded curve), yielding $\hat\rho^{(1)}(t)$ as in Eq.~\eqref{eq.linear_single} except for the missing decaying part in the exponential. 

The fast damping of the induced dipole moment in benzene results from the small size of this molecule, its correspondingly high flexibility, and its high excitation energy. In larger and more rigid molecules like coronene [Fig.~\ref{fig.1}b), inset], the decay is slower, occurring over tens of femtoseconds [Fig.~\ref{fig.1}a), middle panel]. Here, the laser frequency is set to 3.7~eV [Fig.~\ref{fig.1}b), dashed red curve], close to the absorption onset of the molecule. An initial transient polarization during irradiation produces a maximum of the envelope at 8~fs; this is a dispersive rather than absorptive feature, associated with the real part of the polarizability. 
The subsequent decay is not monotonic as for benzene, but superimposed with a beating pattern that is missing in the STE calculation.
In the latter scenario, such a nuclear-motion-induced beating is not expected as the laser and consequently the induced nuclear motion is extremely weak: The excited-state electron dynamics in this approach are predominantly mediated by zero-point energy, not by induced wavepacket motion.
On a technical note, the dipole moment statistically converges much faster for coronene than for benzene (Sec.~S4 and Fig.~S5), \textit{i.e.}, fewer trajectories are required for the former molecule, likely owing to its bigger size and larger rigidity. Symmetry is another important aspect: Both benzene and coronene belong to the $D_{6h}$ point group and are thus highly symmetric. As a consequence, neither of them showcases the common scenario of disordered samples.
In such a case, much less selection rules are in effect~\cite{harris1989,cocchi+2014jpca}, leading to richer vibronic dynamics involving many bright electronic states and coupled vibrations. As an example, only 2 out of 30 vibrational modes in benzene are totally symmetric~\cite{wilson1934pr} and, therefore, allowed to couple to electronic excitations within the Franck-Condon (FC) approximation. In a molecule without symmetries, such constraints are absent. 

Motivated by these considerations, we examine a less symmetric conjugated molecule obtained by isoelectronically replacing one CH group in coronene by an N atom [Fig.~\ref{fig.1}b), inset]. The laser frequency is set to 3.9~eV, slightly above the absorption onset [Fig.~\ref{fig.1}b), dashed blue curve]. Already in the single-trajectory dynamics, a persistent beating pattern appears in the induced dipole moment [Fig.~\ref{fig.1}a), bottom]. This is a fingerprint of the large number of optically-active states participating in the dynamics, which still predominantly occur within the linear regime, resulting from a density matrix of the form
\begin{align}\label{eq.linear_multi}
    \hat\rho^{(1)}(t) \sim \sum_{m=1}^M\rho_{0m}\exp(i\omega_{0m}t)|0\rangle\langle m|,
\end{align}
involving a total of $M$ ES within the frequency band of the laser. The coefficients $\rho_{0m}$ depend on the coupling strength between the electric field and the $|0\rangle\rightarrow|m\rangle$ transition. As for coronene, the ensemble-averaged dipole decays over time, supplying the exponents in Eq.~\eqref{eq.linear_multi} with a real-valued part, $i\omega_{0m}t\rightarrow i\omega_{0m}t-t/T_{0m}$. 
From the induced dipole moments, we are able to calculate within a limited frequency band the optical absorption spectrum (see Supplementary Material, Sec.~S1), for which related nuclear-ensemble-based methods were previously employed~\cite{barbatti+2010pccp, crespo-oteroBarbatti2012tcacc, lively+2021jpcl}. 
Their success in predicting linewidths validates our results, as these widths are closely related to the polarization decay.

In the spectrum of coronene, the so-called $\beta$-band~\cite{clar1964pah}, corresponding to the bright $1^1E_{1u}\leftarrow 1^1A_{1g}$ excitation, exhibits some structure at the low-energy end and a shoulder at the high-energy side [Fig.~\ref{fig.1}b)], in agreement with experiments~\cite{ohno+1972bull, cataldo+2011full}. Comparing to linear-response calculations of the molecule in equilibrium (Fig.~S1), the peak is red-shifted as a sign of FC-type vibronic coupling, corresponding to a non-vanishing curvature in the FC region, \textit{i.e.}, the vertical projection of the ground-state distribution onto ES surfaces.
The finite oscillator strength at 3.1~eV is a consequence of Herzberg-Teller (HT) vibronic coupling. Like the bright $1^1E_{1u}\leftarrow 1^1A_{1g}$ excitation, the $1^1B_{2u}\leftarrow 1^1A_{1g}$ one, corresponding to the so-called $p$-band~\cite{clar1964pah}, arises from transitions between the double-degenerate highest occupied and lowest unoccupied orbitals. In the FC approximation, it is strictly forbidden, as the two equivalent configurations involving degenerate frontier orbitals are superposed destructively. However, symmetry-breaking fluctuations of the nuclear positions due to ZPE give rise to a non-zero transition dipole moment: the excitation borrows intensity from $1^1E_{1u}\leftarrow 1^1A_{1g}$, enabled by their energetic proximity.

Compared to the pristine counterpart, N-coronene absorbs less in the considered energy window [Fig.~\ref{fig.1}b)]. The main peak exhibits several satellites due to optical activation of dark states by substitution-related symmetry lowering (Fig.~S1). While the $p$-band is no longer visible, likely due to negligible laser power at corresponding frequencies, additional dark states emerge at 3.3~eV, draining oscillator strength. As the laser spectrum is not centered on the peak, induced dipolar oscillations are rather weak [Fig.~\ref{fig.1}a)]. This leads to a higher number of trajectories to statistically converge the dipole moment (Fig.~S5); the higher amount of bright states and FC-coupled vibrational modes plays a role, too.

\begin{figure*}
    \centering
    \includegraphics[width=.95\textwidth]{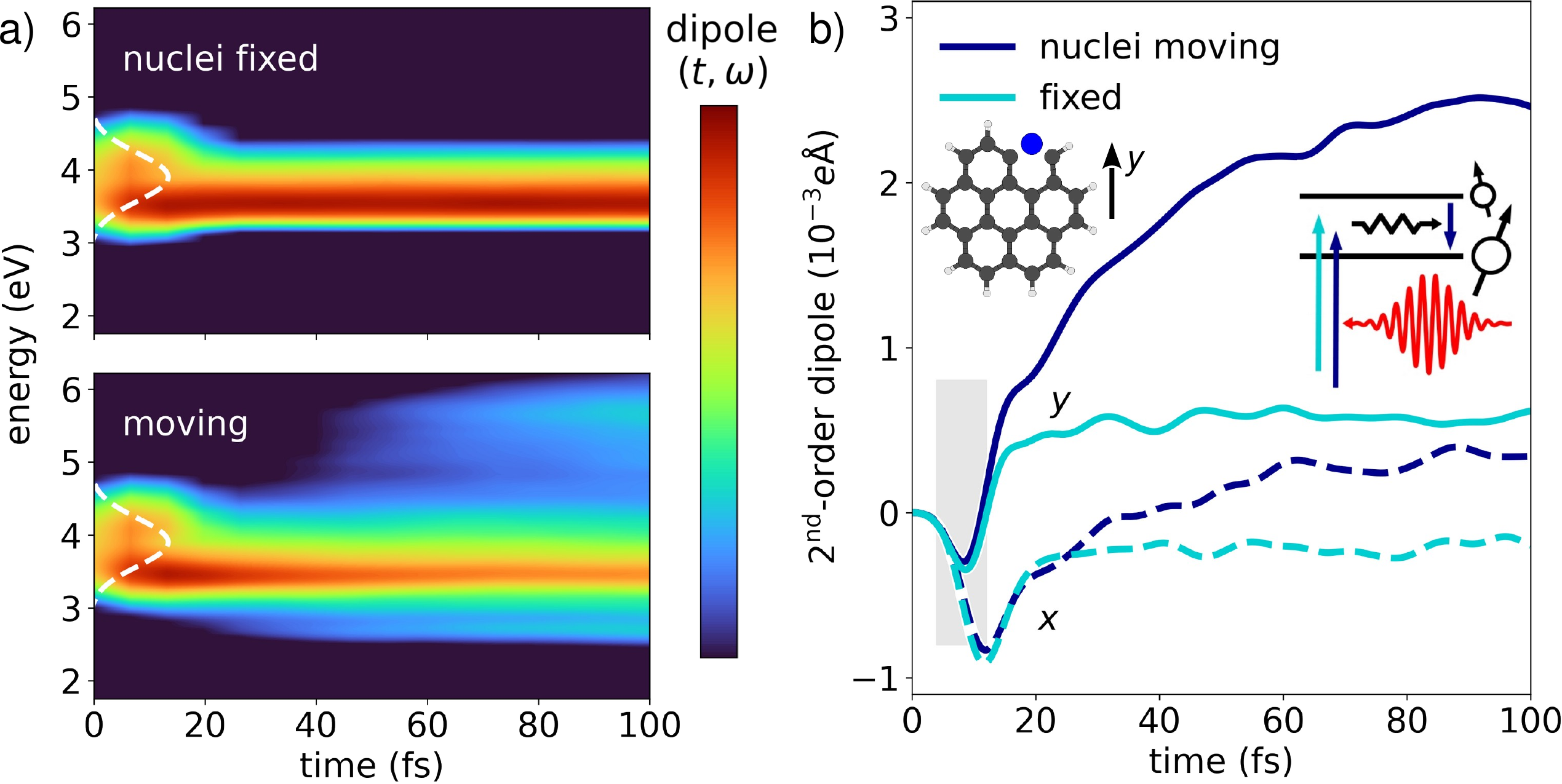}
    \caption{a) Time- and frequency-resolved dipole moment of N-coronene with the laser spectrum shown by white dashed lines. b) Slow components of the dipole moment resolved in $x$ and $y$ (dashed and solid lines, respectively, inset) with the grey bar marking the interval of laser irradiation. Right inset: schematic illustration of the time evolution, involving two excited states with different dipole moments.}
    \label{fig.2}
\end{figure*}

We perform additional calculations with the same nuclear ensemble, but keeping the atoms fixed. This scenario can be expected to yield results equivalent to those from snapshot-based nuclear-ensemble approach based on linear-response TDDFT~\cite{barbatti+2010pccp, crespo-oteroBarbatti2012tcacc}, given the good agreement between the linear absorption spectra predicted by real-time propagation and the Casida equation formalism (Fig.~S1).
Comparing the fully dynamical to the snapshot results, we can assess the role of nuclear momentum in the coupled dynamics. We highlight two differences in the resulting spectra [Fig.~\ref{fig.1}b)]: 
(i) Main absorption features are smeared out by
the nuclear motion; details in the spectra arise from long-term time evolution, which here is damped by moving ions. 
(ii) Nuclear motion leads to a redistribution of the oscillator strength at 3.5~eV to lower energy. In time domain, this is reflected in stronger mid-term dipolar oscillations (40-60~fs window, not shown). We attribute both (i) and (ii) to non-adiabatic coupling: Population is transferred from weak transitions at 3.8~eV to bright ones at 3.6~eV, and finally to the dark states below the onset. Such processes are mediated by nuclear momentum and thus missing in static-nuclei calculations. 

In the field of non-adiabatic molecular dynamics, the term ``coherence'' often evokes associations with the localization of nuclear wavefunctions on different potential-energy surfaces (PES). ``Decoherence'' is thus understood mainly as nuclear wavepackets travelling towards different regions in configurational space. This is manifested, \textit{e.g.}, in the definition of coherence indicators based on integrals over absolute values of nuclear wavefunctions~\cite{min+2017jpcl}, neglecting the complex phase. In this sense, there is no decoherence in Ehrenfest dynamics. However, writing the coherence between the ground state $g$ and an ES $e$, with respective wavepackets $\chi_g$ and $\chi_e$, as
\begin{align}
    \rho_{ge}(t) = \int\text d{\cal Q}\,\chi_g^*({\cal Q},t)\chi_e({\cal Q},t),
\end{align}
it is clear that it does not solely decrease due to the divergence of wavepackets, but also due to internal dephasing between $\chi_g$ and $\chi_e$. In the limiting case of an instantaneous optical excitation, $e\leftarrow g$, part of $\chi_g$ is elevated to an ES surface to form $\chi_e$, where it no longer corresponds to an energy eigenstate and thus undergoes a non-trivial and ${\cal Q}$-dependent phase evolution, while $\chi_g({\cal Q},t) = \chi_g({\cal Q},0)\exp(-i\omega_gt/2)$. As a result, $\rho_{ge}$ can quickly diminish even while $\chi_e$ has not yet left the FC region. This contribution to the loss of coherence is captured by MTE through the distribution of electronic transition frequencies. The good agreement of the predicted initial dipole dynamics with exact results recently shown by Albareda \textit{et al.} \cite{albareda2021jctc} indicates that the internal dephasing occurs significantly faster than the departure of $\chi_e$ from the FC region of the ES surface. Overlap revivals due to rephasing of $\chi_g$ and $\chi_e$ - the cause of FC replica in optical spectra~\cite{heller1976jcp} - are, however, not captured, as the coherence decay is irreversible due to the non-quantized distribution of transition frequencies. In larger systems, such recurrences become increasingly unlikely, as they have to take place in all vibrational modes simultaneously; a vanishing overlap for a single mode renders the total overlap zero.

\subsection{Electron Dynamics: Nonlinear Regime}

Populations of many-body states are diagonal elements of the density matrix and thus can be reached only through two-photon absorption. Consequently, they come into play in second-order processes, buried underneath the dominant linear response. Such populations are not directly accessible in the adopted real-time implementation. 
However, indications can be drawn from the time-resolved fluorescence, hereby calculated performing short-time Fourier transforms of individual dipole moments~\cite{kuda+2020jpca} (Sec.~S1). The ensemble-averaged result remains constant after some initial transient polarization during laser irradiation if nuclei are frozen [Fig.~\ref{fig.2}a), top panel]: no population is transferred to other states, as anticipated. By enabling nuclear motion, other frequency components are mixed-in over time at the expense of the optically targeted state, mainly resulting from an energetically lower state at 2.7~eV [Fig.~\ref{fig.2}a), bottom panel]. However, higher-lying states participate, too. 
We note that this population transfer is mediated by the zero-point energy, not by the induced nuclear motion. Indeed, the latter is very weak in the employed formalism, since only a weak pulse is applied. This can be expected to work well only for systems that do not reorganize significantly upon excitation, \textit{i.e.}, whose targeted ES surfaces do not differ qualitatively from the ground-state one and support bound states.
 
 More aspects of the ES dynamics can be illuminated using arguments based on perturbation theory. ES populations and coherences, $\hat\rho^{(2)}(t) ~\sim\rho_{mn}\exp(i\omega_{mn}t)|m\rangle\langle n|$, 
%
%
where $|m\rangle$ and $|n\rangle$ are both ES, are part of the second-order response. Coherences ($m\neq n$) tend to have oscillation periods $\omega_{mn}$, similar to those of interatomic vibrations, and have been conjectured to play a key role in energy transport in certain photosynthetic complexes~\cite{brixner+2005nature, engel+2007nature, lee+2007science, collini+2010nature, pani+2010proceedings, hildner+2013sci}. The second-order response additionally contains second-harmonic and Raman terms, both involving the ground state (Fig.~S3). All these superposed processes can be partially separated using phase-cycling~\cite{seidner+1995jcp} or low-pass filtering (Sec.~S3), exploiting the fact that second-harmonic processes, as well as linear ones, entail a distinctively fast time evolution.

We now investigate the ES dynamics occurring in N-coronene after excitation with a pulse centered at 3.9~eV. Due to the lack of inversion invariance in this system, second-order contributions tend to be dipolar in character, which is not possible in centrosymmetric compounds like coronene. In the N-substituted molecule we indeed find a non-zero second-order dipole after irradiation [blue curves in Fig.~\ref{fig.2}b)]. For a brief period after excitation, coherence between ES is maintained, as evident in the initial dip, corresponding to a single oscillation cycle. Afterwards, this evolution is taken over by an incoherent buildup of dipole moment on a timescale of $\sim$100~fs, associated with population transfer between states. Unsurprisingly, this effect is absent in fixed-nuclei simulations [Fig.~\ref{fig.2}b), turquoise curves]. The dipole moment after 100~fs differs significantly by magnitude and orientation in the two scenarios [Fig.~\ref{fig.2}b), right inset]: with enabled nuclear dynamics, it points towards the N atom, indicating charge transfer to its site after excitation of the delocalized $\pi$-conjugated network.
The dipole moment does not always reflect the ES dynamics as unambiguously as in this case. In general, one can resort to other observables directly related to the electron density, such as partial charges or higher multipole moments. Compared to other methods for non-adiabatic dynamics, the focus is thus shifted away from PES towards a real-space representation based on the charge density. Further conclusions can be drawn from direct simulations of third-order spectroscopy, which is straightforward with RT-TDDFT or related methods~\cite{umberto+2013cpc,bonafe+2018jpcl, hernandez+2019jpca, krumland+2020jcp, herperger+2021jpca}.

\begin{figure}
    \centering
    \includegraphics[width=0.475\textwidth]{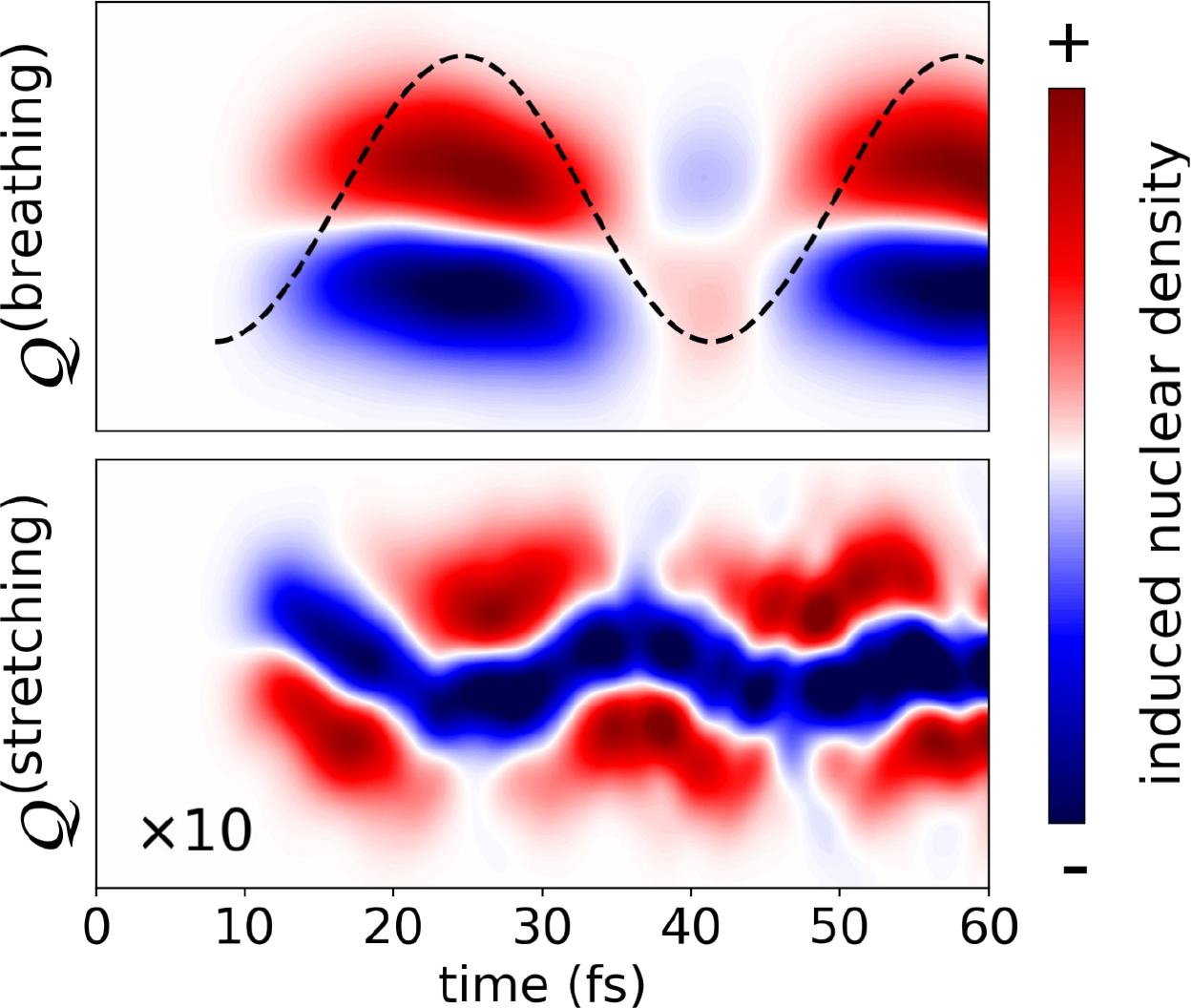}
    \caption{Wave-packet motion caused by electronic excitation of benzene, achieved by applying a laser pulse with a frequency of about 6.9~eV. Positive (red) and negative (blue) areas indicate increased and decreased nuclear probability density with respect to the ground state, respectively. The top panel shows the dominant totally symmetric breathing mode, the bottom one a less symmetric, yet Herzberg-Teller active stretching mode. The dashed line in the top panel is a sine function with a frequency of 1000~cm$^{-1}$, matching that of the breathing mode. }
    \label{fig.3}
\end{figure}

\subsection{Nuclear Dynamics}

Finally, we analyze vibrational coherence established by electronic excitation. This corresponds to harmonic wavepacket motion after instantaneous elevation to a harmonic ES PES and is characterized by an essentially classical time evolution~\cite{schleich2011quantum}. Thus, such motion does not need to be viewed as arising from quantum interference, even though the term ``coherence'' tends to invoke such associations~\cite{miller2012jcp}.
In the following, we consider benzene as an example. As mentioned, this molecule has only two totally symmetric modes that can couple to electronic excitations within the FC approximation. One of them is the breathing mode around 1000~cm$^{-1}$, while the other one is a C-H stretching mode, which, however, is not effectively stimulated by the $\pi-\pi^*$ transition at 6.9~eV. In STE, which yields only FC-type dynamics, nuclear motion is thus induced predominantly in the breathing mode. For MTE, we visualize the induced nuclear density (Fig.~\ref{fig.3}) which is the probability distribution of the nuclei relative to the ground state (Sec.~S1). For the breathing mode (upper panel), we indeed find an approximately harmonic time evolution. A corresponding classical trajectory is given by the superposed sine function (dashed line). At the maxima, there are adjacent regions of density accumulation and depletion, reflecting a wavepacket having departed the FC region of the ES surface, leaving behind a hole in the ground state; minima mark instances of return. The slightly tilted shape of high-density regions as well as the density not being strictly zero at the minima is the result of using a realistically shaped laser pulse instead of an instantaneous promotion to the ES. We note again that due to its mean-field nature, MTE does not actually include wavepacket splitting, but rather describes the dynamics through a single, averaged nuclear wavepacket. However, these different descriptions appear qualitatively similarly in the nuclear density if a ground-state reference is subtracted, as done here.

In MTE, modes without the complete symmetry of the structure gain energy, too, like the C=C stretching mode around 1600~cm$^{-1}$, which transforms as $e_{2g}$, thus breaking the hexagonal symmetry of the molecule. Vibrations of this representation activate otherwise dark $n^1B_{2u}\leftarrow 1^1A_{1g}$ excitations through HT-type coupling~\cite{li+2010pccp}. This type of post-FC effect can be interpreted as a fingerprint of electron-nuclear correlation, requiring a ${\cal Q}$-dependent transition-dipole moment and thus parametrically ${\cal Q}$-dependent electronic wavefunctions, with ${\cal Q}$ being normal-mode coordinates (Sec.~S1). In this case, the intuitive picture of a Gaussian wavepacket prepared and classically moving on an ES PES is invalid, and the observed nuclear dynamics are qualitatively different (Fig.~\ref{fig.3}, bottom). Particularly, perfect overlap between the wavepacket and the hole left behind in the ground state is never recovered. There is a stationary redistribution of nuclear density from the center to the outskirts of the oscillator, plus a superposed harmonic oscillation which is damped over time. 

\section{Summary and Conclusions}

In summary, we have investigated the combination of RT-TDDFT+Ehrenfest and quantum sampling of initial conditions for \textit{ab initio} simulations of laser-induced ultrafast coherent dynamics applied to conjugated molecules. Contrary to the single-trajectory version, this approach naturally includes electronic dephasing effects without empirical parameters and is capable of describing transitions between close-lying ES. Furthermore, it can be employed to determine laser-induced nuclear dynamics such as coherent wavepacket motion of FC-coupled vibrational modes, or non-classical nuclear dynamics associated with purely HT-active excitations. While the considered molecules are medium-sized, the computational efficiency and scalability of RT-TDDFT+Ehrenfest favor application to larger systems. 

We assume the validity of the proposed scheme to increase with system size for several reasons: (i) Duschinsky rotations and vibrational frequency shifts tend to be much smaller; (ii) a revival of electronic coherence due to the wavepacket returning to the FC region -- which MTE does not seem to capture properly -- becomes unlikely; (iii) the excitation-induced electron-density perturbation becomes smaller in relation to the total density, presumably enhancing the validity of the adiabatic approximation assumed for the dynamical exchange-correlation potential. This rationale holds mainly for electrons in delocalized orbitals; for strongly localized ones, the excitation-induced density in relation to the total density is more akin to smaller molecules.
The optimal trade-off between accuracy, computational efficiency, and insight into the involved physical processes offered by the proposed method unfolds bright perspectives for \textit{ab initio} simulations of ultrafast coherent spectroscopies as a key complement to corresponding experiments.
In future work, extensions to various flavors of nonlinear~\cite{cocchi2014,guandalini2021} and multidimensional spectroscopies~\cite{desi+19zna,coll21jpcc} are foreseen. While the statistical aspects  naturally lead towards machine learning~\cite{chen+2020jpcl, xue+2020jpca}, methodological progress can be achieved by coupling trajectories during the time evolution, thereby inducing
wavepacket splitting~\cite{min+2015prl, curc+2018epjb, gossel+2018jctc}.

\vspace{0.5 cm}
\section*{Acknowledgements}
We are thankful to Michele Guerrini, Katherine R. Herperger, and Mariana Rossi for fruitful discussions, and to Ralph Ernstorfer for posing the question that stimulated this research.
This work was funded by the German Research Foundation (DFG), project number 182087777 -- CRC 951, by the German Federal Ministry of Education and Research (Professorinnenprogramm III), and by the State of Lower Saxony (Professorinnen für Niedersachsen).
Computational resources were provided by the North-German Supercomputing
Alliance (HLRN), project bep00076.


\begin{thebibliography}{69}%
\makeatletter
\providecommand \@ifxundefined [1]{%
 \@ifx{#1\undefined}
}%
\providecommand \@ifnum [1]{%
 \ifnum #1\expandafter \@firstoftwo
 \else \expandafter \@secondoftwo
 \fi
}%
\providecommand \@ifx [1]{%
 \ifx #1\expandafter \@firstoftwo
 \else \expandafter \@secondoftwo
 \fi
}%
\providecommand \natexlab [1]{#1}%
\providecommand \enquote  [1]{``#1''}%
\providecommand \bibnamefont  [1]{#1}%
\providecommand \bibfnamefont [1]{#1}%
\providecommand \citenamefont [1]{#1}%
\providecommand \href@noop [0]{\@secondoftwo}%
\providecommand \href [0]{\begingroup \@sanitize@url \@href}%
\providecommand \@href[1]{\@@startlink{#1}\@@href}%
\providecommand \@@href[1]{\endgroup#1\@@endlink}%
\providecommand \@sanitize@url [0]{\catcode `\\12\catcode `\$12\catcode
  `\&12\catcode `\#12\catcode `\^12\catcode `\_12\catcode `\%12\relax}%
\providecommand \@@startlink[1]{}%
\providecommand \@@endlink[0]{}%
\providecommand \url  [0]{\begingroup\@sanitize@url \@url }%
\providecommand \@url [1]{\endgroup\@href {#1}{\urlprefix }}%
\providecommand \urlprefix  [0]{URL }%
\providecommand \Eprint [0]{\href }%
\providecommand \doibase [0]{https://doi.org/}%
\providecommand \selectlanguage [0]{\@gobble}%
\providecommand \bibinfo  [0]{\@secondoftwo}%
\providecommand \bibfield  [0]{\@secondoftwo}%
\providecommand \translation [1]{[#1]}%
\providecommand \BibitemOpen [0]{}%
\providecommand \bibitemStop [0]{}%
\providecommand \bibitemNoStop [0]{.\EOS\space}%
\providecommand \EOS [0]{\spacefactor3000\relax}%
\providecommand \BibitemShut  [1]{\csname bibitem#1\endcsname}%
\let\auto@bib@innerbib\@empty
\bibitem [{\citenamefont {Mukamel}(2000)}]{mukamel2000arpc}%
  \BibitemOpen
  \bibfield  {author} {\bibinfo {author} {\bibfnamefont {S.}~\bibnamefont
  {Mukamel}},\ }\bibfield  {title} {\bibinfo {title} {Multidimensional
  femtosecond correlation spectroscopies of electronic and vibrational
  excitations},\ }\href@noop {} {\bibfield  {journal} {\bibinfo  {journal}
  {Annu.~Rev.~Phys.~Chem.~}\ }\textbf {\bibinfo {volume} {51}},\ \bibinfo
  {pages} {691} (\bibinfo {year} {2000})}\BibitemShut {NoStop}%
\bibitem [{\citenamefont {Jonas}(2003)}]{jonas2003arpc}%
  \BibitemOpen
  \bibfield  {author} {\bibinfo {author} {\bibfnamefont {D.~M.}\ \bibnamefont
  {Jonas}},\ }\bibfield  {title} {\bibinfo {title} {Two-dimensional femtosecond
  spectroscopy},\ }\href@noop {} {\bibfield  {journal} {\bibinfo  {journal}
  {Annu.~Rev.~Phys.~Chem.~}\ }\textbf {\bibinfo {volume} {54}},\ \bibinfo
  {pages} {425} (\bibinfo {year} {2003})}\BibitemShut {NoStop}%
\bibitem [{\citenamefont {Cho}(2008)}]{cho2008cr}%
  \BibitemOpen
  \bibfield  {author} {\bibinfo {author} {\bibfnamefont {M.}~\bibnamefont
  {Cho}},\ }\bibfield  {title} {\bibinfo {title} {Coherent two-dimensional
  optical spectroscopy},\ }\href {https://doi.org/10.1021/cr078377b} {\bibfield
   {journal} {\bibinfo  {journal} {Chem.~Rev.~}\ }\textbf {\bibinfo {volume}
  {108}},\ \bibinfo {pages} {1331} (\bibinfo {year} {2008})},\ \bibinfo {note}
  {pMID: 18363410},\ \Eprint
  {https://arxiv.org/abs/https://doi.org/10.1021/cr078377b}
  {https://doi.org/10.1021/cr078377b} \BibitemShut {NoStop}%
\bibitem [{\citenamefont {Hamm}\ and\ \citenamefont
  {Zanni}(2011)}]{hamm+zanni2011}%
  \BibitemOpen
  \bibfield  {author} {\bibinfo {author} {\bibfnamefont {P.}~\bibnamefont
  {Hamm}}\ and\ \bibinfo {author} {\bibfnamefont {M.}~\bibnamefont {Zanni}},\
  }\href {https://doi.org/10.1017/CBO9780511675935} {\emph {\bibinfo {title}
  {Concepts and Methods of 2D Infrared Spectroscopy}}}\ (\bibinfo  {publisher}
  {Cambridge University Press},\ \bibinfo {year} {2011})\BibitemShut {NoStop}%
\bibitem [{\citenamefont {Collini}(2021)}]{coll21jpcc}%
  \BibitemOpen
  \bibfield  {author} {\bibinfo {author} {\bibfnamefont {E.}~\bibnamefont
  {Collini}},\ }\bibfield  {title} {\bibinfo {title} {2d electronic
  spectroscopic techniques for quantum technology applications},\ }\href
  {https://doi.org/10.1021/acs.jpcc.1c02693} {\bibfield  {journal} {\bibinfo
  {journal} {J.~Phys.~Chem.~C}\ }\textbf {\bibinfo {volume} {125}},\ \bibinfo
  {pages} {13096} (\bibinfo {year} {2021})},\ \Eprint
  {https://arxiv.org/abs/https://doi.org/10.1021/acs.jpcc.1c02693}
  {https://doi.org/10.1021/acs.jpcc.1c02693} \BibitemShut {NoStop}%
\bibitem [{\citenamefont {Conti}\ \emph {et~al.}(2020)\citenamefont {Conti},
  \citenamefont {Cerullo}, \citenamefont {Nenov},\ and\ \citenamefont
  {Garavelli}}]{conti+20jacs}%
  \BibitemOpen
  \bibfield  {author} {\bibinfo {author} {\bibfnamefont {I.}~\bibnamefont
  {Conti}}, \bibinfo {author} {\bibfnamefont {G.}~\bibnamefont {Cerullo}},
  \bibinfo {author} {\bibfnamefont {A.}~\bibnamefont {Nenov}},\ and\ \bibinfo
  {author} {\bibfnamefont {M.}~\bibnamefont {Garavelli}},\ }\bibfield  {title}
  {\bibinfo {title} {Ultrafast spectroscopy of photoactive molecular systems
  from first principles: Where we stand today and where we are going},\
  }\href@noop {} {\bibfield  {journal} {\bibinfo  {journal}
  {J.~Am.~Chem.~Soc.~}\ }\textbf {\bibinfo {volume} {142}},\ \bibinfo {pages}
  {16117} (\bibinfo {year} {2020})}\BibitemShut {NoStop}%
\bibitem [{\citenamefont {Curchod}\ and\ \citenamefont
  {Martínez}(2018)}]{curc+mart2018cr}%
  \BibitemOpen
  \bibfield  {author} {\bibinfo {author} {\bibfnamefont {B.~F.~E.}\
  \bibnamefont {Curchod}}\ and\ \bibinfo {author} {\bibfnamefont {T.~J.}\
  \bibnamefont {Martínez}},\ }\bibfield  {title} {\bibinfo {title} {Ab initio
  nonadiabatic quantum molecular dynamics},\ }\href
  {https://doi.org/10.1021/acs.chemrev.7b00423} {\bibfield  {journal} {\bibinfo
   {journal} {Chem.~Rev.~}\ }\textbf {\bibinfo {volume} {118}},\ \bibinfo
  {pages} {3305} (\bibinfo {year} {2018})},\ \bibinfo {note} {pMID: 29465231},\
  \Eprint {https://arxiv.org/abs/https://doi.org/10.1021/acs.chemrev.7b00423}
  {https://doi.org/10.1021/acs.chemrev.7b00423} \BibitemShut {NoStop}%
\bibitem [{\citenamefont {C.~Tully}(1998)}]{tully1998faraday}%
  \BibitemOpen
  \bibfield  {author} {\bibinfo {author} {\bibfnamefont {J.}~\bibnamefont
  {C.~Tully}},\ }\bibfield  {title} {\bibinfo {title} {Mixed
  quantum–classical dynamics},\ }\href {https://doi.org/10.1039/A801824C}
  {\bibfield  {journal} {\bibinfo  {journal} {Faraday Discuss.}\ }\textbf
  {\bibinfo {volume} {110}},\ \bibinfo {pages} {407} (\bibinfo {year}
  {1998})}\BibitemShut {NoStop}%
\bibitem [{\citenamefont {Heller}(1976{\natexlab{a}})}]{heller1976jcp_tdscf}%
  \BibitemOpen
  \bibfield  {author} {\bibinfo {author} {\bibfnamefont {E.~J.}\ \bibnamefont
  {Heller}},\ }\bibfield  {title} {\bibinfo {title} {Time dependent variational
  approach to semiclassical dynamics},\ }\href
  {https://doi.org/10.1063/1.431911} {\bibfield  {journal} {\bibinfo  {journal}
  {J.~Chem.~Phys.~}\ }\textbf {\bibinfo {volume} {64}},\ \bibinfo {pages} {63}
  (\bibinfo {year} {1976}{\natexlab{a}})},\ \Eprint
  {https://arxiv.org/abs/https://doi.org/10.1063/1.431911}
  {https://doi.org/10.1063/1.431911} \BibitemShut {NoStop}%
\bibitem [{\citenamefont {Rozzi}\ \emph {et~al.}(2013)\citenamefont {Rozzi},
  \citenamefont {Falke}, \citenamefont {Spallanzani}, \citenamefont {Rubio},
  \citenamefont {Molinari}, \citenamefont {Brida}, \citenamefont {Maiuri},
  \citenamefont {Cerullo}, \citenamefont {Schramm}, \citenamefont
  {Christoffers},\ and\ \citenamefont {Lienau}}]{rozzi2013natcom}%
  \BibitemOpen
  \bibfield  {author} {\bibinfo {author} {\bibfnamefont {C.~A.}\ \bibnamefont
  {Rozzi}}, \bibinfo {author} {\bibfnamefont {S.~M.}\ \bibnamefont {Falke}},
  \bibinfo {author} {\bibfnamefont {N.}~\bibnamefont {Spallanzani}}, \bibinfo
  {author} {\bibfnamefont {A.}~\bibnamefont {Rubio}}, \bibinfo {author}
  {\bibfnamefont {E.}~\bibnamefont {Molinari}}, \bibinfo {author}
  {\bibfnamefont {D.}~\bibnamefont {Brida}}, \bibinfo {author} {\bibfnamefont
  {M.}~\bibnamefont {Maiuri}}, \bibinfo {author} {\bibfnamefont
  {G.}~\bibnamefont {Cerullo}}, \bibinfo {author} {\bibfnamefont
  {H.}~\bibnamefont {Schramm}}, \bibinfo {author} {\bibfnamefont
  {J.}~\bibnamefont {Christoffers}},\ and\ \bibinfo {author} {\bibfnamefont
  {C.}~\bibnamefont {Lienau}},\ }\bibfield  {title} {\bibinfo {title} {Quantum
  coherence controls the charge separation in a prototypical artificial
  light-harvesting system},\ }\href {https://doi.org/10.1038/ncomms2603}
  {\bibfield  {journal} {\bibinfo  {journal} {Nature Commun.}\ }\textbf
  {\bibinfo {volume} {4}},\ \bibinfo {pages} {1602} (\bibinfo {year}
  {2013})}\BibitemShut {NoStop}%
\bibitem [{\citenamefont {Falke}\ \emph {et~al.}(2014)\citenamefont {Falke},
  \citenamefont {Rozzi}, \citenamefont {Brida}, \citenamefont {Maiuri},
  \citenamefont {Amato}, \citenamefont {Sommer}, \citenamefont {Sio},
  \citenamefont {Rubio}, \citenamefont {Cerullo}, \citenamefont {Molinari},\
  and\ \citenamefont {Lienau}}]{falke+2014sci}%
  \BibitemOpen
  \bibfield  {author} {\bibinfo {author} {\bibfnamefont {S.~M.}\ \bibnamefont
  {Falke}}, \bibinfo {author} {\bibfnamefont {C.~A.}\ \bibnamefont {Rozzi}},
  \bibinfo {author} {\bibfnamefont {D.}~\bibnamefont {Brida}}, \bibinfo
  {author} {\bibfnamefont {M.}~\bibnamefont {Maiuri}}, \bibinfo {author}
  {\bibfnamefont {M.}~\bibnamefont {Amato}}, \bibinfo {author} {\bibfnamefont
  {E.}~\bibnamefont {Sommer}}, \bibinfo {author} {\bibfnamefont {A.~D.}\
  \bibnamefont {Sio}}, \bibinfo {author} {\bibfnamefont {A.}~\bibnamefont
  {Rubio}}, \bibinfo {author} {\bibfnamefont {G.}~\bibnamefont {Cerullo}},
  \bibinfo {author} {\bibfnamefont {E.}~\bibnamefont {Molinari}},\ and\
  \bibinfo {author} {\bibfnamefont {C.}~\bibnamefont {Lienau}},\ }\bibfield
  {title} {\bibinfo {title} {Coherent ultrafast charge transfer in an organic
  photovoltaic blend},\ }\href {https://doi.org/10.1126/science.1249771}
  {\bibfield  {journal} {\bibinfo  {journal} {Science}\ }\textbf {\bibinfo
  {volume} {344}},\ \bibinfo {pages} {1001} (\bibinfo {year}
  {2014})}\BibitemShut {NoStop}%
\bibitem [{\citenamefont {Rozzi}\ \emph {et~al.}(2017)\citenamefont {Rozzi},
  \citenamefont {Troiani},\ and\ \citenamefont {Tavernelli}}]{rozzi+2017jpcm}%
  \BibitemOpen
  \bibfield  {author} {\bibinfo {author} {\bibfnamefont {C.~A.}\ \bibnamefont
  {Rozzi}}, \bibinfo {author} {\bibfnamefont {F.}~\bibnamefont {Troiani}},\
  and\ \bibinfo {author} {\bibfnamefont {I.}~\bibnamefont {Tavernelli}},\
  }\bibfield  {title} {\bibinfo {title} {Quantum modeling of ultrafast
  photoinduced charge separation},\ }\href
  {https://doi.org/10.1088/1361-648x/aa948a} {\bibfield  {journal} {\bibinfo
  {journal} {J.~Phys.:~Condens.~Matter.~}\ }\textbf {\bibinfo {volume} {30}},\
  \bibinfo {pages} {013002} (\bibinfo {year} {2017})}\BibitemShut {NoStop}%
\bibitem [{\citenamefont {Zhang}\ \emph {et~al.}(2017)\citenamefont {Zhang},
  \citenamefont {Hong}, \citenamefont {Lian}, \citenamefont {Ma}, \citenamefont
  {Xu}, \citenamefont {Zhou}, \citenamefont {Fu}, \citenamefont {Liu},\ and\
  \citenamefont {Meng}}]{zhang+2017as}%
  \BibitemOpen
  \bibfield  {author} {\bibinfo {author} {\bibfnamefont {J.}~\bibnamefont
  {Zhang}}, \bibinfo {author} {\bibfnamefont {H.}~\bibnamefont {Hong}},
  \bibinfo {author} {\bibfnamefont {C.}~\bibnamefont {Lian}}, \bibinfo {author}
  {\bibfnamefont {W.}~\bibnamefont {Ma}}, \bibinfo {author} {\bibfnamefont
  {X.}~\bibnamefont {Xu}}, \bibinfo {author} {\bibfnamefont {X.}~\bibnamefont
  {Zhou}}, \bibinfo {author} {\bibfnamefont {H.}~\bibnamefont {Fu}}, \bibinfo
  {author} {\bibfnamefont {K.}~\bibnamefont {Liu}},\ and\ \bibinfo {author}
  {\bibfnamefont {S.}~\bibnamefont {Meng}},\ }\bibfield  {title} {\bibinfo
  {title} {Interlayer-state-coupling dependent ultrafast charge transfer in
  mos2/ws2 bilayers},\ }\href
  {https://doi.org/https://doi.org/10.1002/advs.201700086} {\bibfield
  {journal} {\bibinfo  {journal} {Adv.~Sci.}\ }\textbf {\bibinfo {volume}
  {4}},\ \bibinfo {pages} {1700086} (\bibinfo {year} {2017})}\BibitemShut
  {NoStop}%
\bibitem [{\citenamefont {Jacobs}\ \emph {et~al.}(2020)\citenamefont {Jacobs},
  \citenamefont {Krumland}, \citenamefont {Valencia}, \citenamefont {Wang},
  \citenamefont {Rossi},\ and\ \citenamefont {Cocchi}}]{jaco+2020advpx}%
  \BibitemOpen
  \bibfield  {author} {\bibinfo {author} {\bibfnamefont {M.}~\bibnamefont
  {Jacobs}}, \bibinfo {author} {\bibfnamefont {J.}~\bibnamefont {Krumland}},
  \bibinfo {author} {\bibfnamefont {A.~M.}\ \bibnamefont {Valencia}}, \bibinfo
  {author} {\bibfnamefont {H.}~\bibnamefont {Wang}}, \bibinfo {author}
  {\bibfnamefont {M.}~\bibnamefont {Rossi}},\ and\ \bibinfo {author}
  {\bibfnamefont {C.}~\bibnamefont {Cocchi}},\ }\bibfield  {title} {\bibinfo
  {title} {Ultrafast charge transfer and vibronic coupling in a laser-excited
  hybrid inorganic/organic interface},\ }\href
  {https://doi.org/10.1080/23746149.2020.1749883} {\bibfield  {journal}
  {\bibinfo  {journal} {Adv.~Phys.:~X}\ }\textbf {\bibinfo {volume} {5}},\
  \bibinfo {pages} {1749883} (\bibinfo {year} {2020})},\ \Eprint
  {https://arxiv.org/abs/https://doi.org/10.1080/23746149.2020.1749883}
  {https://doi.org/10.1080/23746149.2020.1749883} \BibitemShut {NoStop}%
\bibitem [{\citenamefont {De~Sio}\ and\ \citenamefont
  {Lienau}(2017)}]{desi-lien17pccp}%
  \BibitemOpen
  \bibfield  {author} {\bibinfo {author} {\bibfnamefont {A.}~\bibnamefont
  {De~Sio}}\ and\ \bibinfo {author} {\bibfnamefont {C.}~\bibnamefont
  {Lienau}},\ }\bibfield  {title} {\bibinfo {title} {Vibronic coupling in
  organic semiconductors for photovoltaics},\ }\href@noop {} {\bibfield
  {journal} {\bibinfo  {journal} {Phys.~Chem.~Chem.~Phys.~}\ }\textbf {\bibinfo
  {volume} {19}},\ \bibinfo {pages} {18813} (\bibinfo {year}
  {2017})}\BibitemShut {NoStop}%
\bibitem [{\citenamefont {Makri}\ and\ \citenamefont
  {Miller}(1987)}]{nancy+miller1987}%
  \BibitemOpen
  \bibfield  {author} {\bibinfo {author} {\bibfnamefont {N.}~\bibnamefont
  {Makri}}\ and\ \bibinfo {author} {\bibfnamefont {W.~H.}\ \bibnamefont
  {Miller}},\ }\bibfield  {title} {\bibinfo {title} {Time‐dependent
  self‐consistent field (tdscf) approximation for a reaction coordinate
  coupled to a harmonic bath: Single and multiple configuration treatments},\
  }\href {https://doi.org/10.1063/1.453501} {\bibfield  {journal} {\bibinfo
  {journal} {J.~Chem.~Phys.~}\ }\textbf {\bibinfo {volume} {87}},\ \bibinfo
  {pages} {5781} (\bibinfo {year} {1987})},\ \Eprint
  {https://arxiv.org/abs/https://doi.org/10.1063/1.453501}
  {https://doi.org/10.1063/1.453501} \BibitemShut {NoStop}%
\bibitem [{\citenamefont {Hillery}\ \emph {et~al.}(1984)\citenamefont
  {Hillery}, \citenamefont {O'Connell}, \citenamefont {Scully},\ and\
  \citenamefont {Wigner}}]{hillery+1984pr}%
  \BibitemOpen
  \bibfield  {author} {\bibinfo {author} {\bibfnamefont {M.}~\bibnamefont
  {Hillery}}, \bibinfo {author} {\bibfnamefont {R.}~\bibnamefont {O'Connell}},
  \bibinfo {author} {\bibfnamefont {M.}~\bibnamefont {Scully}},\ and\ \bibinfo
  {author} {\bibfnamefont {E.}~\bibnamefont {Wigner}},\ }\bibfield  {title}
  {\bibinfo {title} {Distribution functions in physics: Fundamentals},\ }\href
  {https://doi.org/https://doi.org/10.1016/0370-1573(84)90160-1} {\bibfield
  {journal} {\bibinfo  {journal} {Phys.~Rep.~}\ }\textbf {\bibinfo {volume}
  {106}},\ \bibinfo {pages} {121} (\bibinfo {year} {1984})}\BibitemShut
  {NoStop}%
\bibitem [{\citenamefont {Ullrich}(2012)}]{ullrich2012oxford}%
  \BibitemOpen
  \bibfield  {author} {\bibinfo {author} {\bibfnamefont {C.~A.}\ \bibnamefont
  {Ullrich}},\ }\href
  {https://doi.org/10.1093/acprof:oso/9780199563029.001.0001} {\emph {\bibinfo
  {title} {Time-Dependent Density-Functional Theory: Concepts and
  Applications}}}\ (\bibinfo  {publisher} {Oxford University Press},\ \bibinfo
  {year} {2012})\BibitemShut {NoStop}%
\bibitem [{\citenamefont {Runge}\ and\ \citenamefont
  {Gross}(1984)}]{rung+1984prl}%
  \BibitemOpen
  \bibfield  {author} {\bibinfo {author} {\bibfnamefont {E.}~\bibnamefont
  {Runge}}\ and\ \bibinfo {author} {\bibfnamefont {E.~K.~U.}\ \bibnamefont
  {Gross}},\ }\bibfield  {title} {\bibinfo {title} {Density-functional theory
  for time-dependent systems},\ }\href@noop {} {\bibfield  {journal} {\bibinfo
  {journal} {Phys. Rev. Lett.}\ }\textbf {\bibinfo {volume} {52}},\ \bibinfo
  {pages} {997} (\bibinfo {year} {1984})}\BibitemShut {NoStop}%
\bibitem [{\citenamefont {Hohenberg}\ and\ \citenamefont
  {Kohn}(1964)}]{hohenbergKohn1964pr}%
  \BibitemOpen
  \bibfield  {author} {\bibinfo {author} {\bibfnamefont {P.}~\bibnamefont
  {Hohenberg}}\ and\ \bibinfo {author} {\bibfnamefont {W.}~\bibnamefont
  {Kohn}},\ }\bibfield  {title} {\bibinfo {title} {Inhomogeneous electron
  gas},\ }\href {https://doi.org/10.1103/PhysRev.136.B864} {\bibfield
  {journal} {\bibinfo  {journal} {Phys.~Rev.~}\ }\textbf {\bibinfo {volume}
  {136}},\ \bibinfo {pages} {B864} (\bibinfo {year} {1964})}\BibitemShut
  {NoStop}%
\bibitem [{\citenamefont {Kohn}\ and\ \citenamefont
  {Sham}(1965)}]{kohnSham1965pr}%
  \BibitemOpen
  \bibfield  {author} {\bibinfo {author} {\bibfnamefont {W.}~\bibnamefont
  {Kohn}}\ and\ \bibinfo {author} {\bibfnamefont {L.~J.}\ \bibnamefont
  {Sham}},\ }\bibfield  {title} {\bibinfo {title} {Self-consistent equations
  including exchange and correlation effects},\ }\href
  {https://doi.org/10.1103/PhysRev.140.A1133} {\bibfield  {journal} {\bibinfo
  {journal} {Phys.~Rev.~}\ }\textbf {\bibinfo {volume} {140}},\ \bibinfo
  {pages} {A1133} (\bibinfo {year} {1965})}\BibitemShut {NoStop}%
\bibitem [{\citenamefont {Maitra}\ \emph {et~al.}(2002)\citenamefont {Maitra},
  \citenamefont {Burke},\ and\ \citenamefont {Woodward}}]{mait+2002prl}%
  \BibitemOpen
  \bibfield  {author} {\bibinfo {author} {\bibfnamefont {N.~T.}\ \bibnamefont
  {Maitra}}, \bibinfo {author} {\bibfnamefont {K.}~\bibnamefont {Burke}},\ and\
  \bibinfo {author} {\bibfnamefont {C.}~\bibnamefont {Woodward}},\ }\bibfield
  {title} {\bibinfo {title} {Memory in time-dependent density functional
  theory},\ }\href {https://doi.org/10.1103/PhysRevLett.89.023002} {\bibfield
  {journal} {\bibinfo  {journal} {Phys.~Rev.~Lett.~}\ }\textbf {\bibinfo
  {volume} {89}},\ \bibinfo {pages} {023002} (\bibinfo {year}
  {2002})}\BibitemShut {NoStop}%
\bibitem [{\citenamefont {Lacombe}\ and\ \citenamefont
  {Maitra}(2020)}]{lacombe2020fd}%
  \BibitemOpen
  \bibfield  {author} {\bibinfo {author} {\bibfnamefont {L.}~\bibnamefont
  {Lacombe}}\ and\ \bibinfo {author} {\bibfnamefont {N.~T.}\ \bibnamefont
  {Maitra}},\ }\bibfield  {title} {\bibinfo {title} {Developing new and
  understanding old approximations in tddft},\ }\href
  {https://doi.org/10.1039/D0FD00049C} {\bibfield  {journal} {\bibinfo
  {journal} {Faraday Discuss.}\ }\textbf {\bibinfo {volume} {224}},\ \bibinfo
  {pages} {382} (\bibinfo {year} {2020})}\BibitemShut {NoStop}%
\bibitem [{\citenamefont {Lacombe}\ and\ \citenamefont
  {Maitra}(2021)}]{lacombe+maitra2021jpcl}%
  \BibitemOpen
  \bibfield  {author} {\bibinfo {author} {\bibfnamefont {L.}~\bibnamefont
  {Lacombe}}\ and\ \bibinfo {author} {\bibfnamefont {N.~T.}\ \bibnamefont
  {Maitra}},\ }\bibfield  {title} {\bibinfo {title} {Minimizing the
  time-dependent density functional error in ehrenfest dynamics},\ }\href
  {https://doi.org/10.1021/acs.jpclett.1c02020} {\bibfield  {journal} {\bibinfo
   {journal} {J.~Phys.~Chem.~Lett.}\ }\textbf {\bibinfo {volume} {12}},\
  \bibinfo {pages} {8554} (\bibinfo {year} {2021})},\ \bibinfo {note} {pMID:
  34464148},\ \Eprint
  {https://arxiv.org/abs/https://doi.org/10.1021/acs.jpclett.1c02020}
  {https://doi.org/10.1021/acs.jpclett.1c02020} \BibitemShut {NoStop}%
\bibitem [{\citenamefont {Andrade}\ \emph {et~al.}(2015)\citenamefont
  {Andrade}, \citenamefont {Strubbe}, \citenamefont {De~Giovannini},
  \citenamefont {Larsen}, \citenamefont {Oliveira}, \citenamefont
  {Alberdi-Rodriguez}, \citenamefont {Varas}, \citenamefont {Theophilou},
  \citenamefont {Helbig}, \citenamefont {Verstraete}, \citenamefont {Stella},
  \citenamefont {Nogueira}, \citenamefont {Aspuru-Guzik}, \citenamefont
  {Castro}, \citenamefont {Marques},\ and\ \citenamefont
  {Rubio}}]{octopus2015}%
  \BibitemOpen
  \bibfield  {author} {\bibinfo {author} {\bibfnamefont {X.}~\bibnamefont
  {Andrade}}, \bibinfo {author} {\bibfnamefont {D.}~\bibnamefont {Strubbe}},
  \bibinfo {author} {\bibfnamefont {U.}~\bibnamefont {De~Giovannini}}, \bibinfo
  {author} {\bibfnamefont {A.~H.}\ \bibnamefont {Larsen}}, \bibinfo {author}
  {\bibfnamefont {M.~J.~T.}\ \bibnamefont {Oliveira}}, \bibinfo {author}
  {\bibfnamefont {J.}~\bibnamefont {Alberdi-Rodriguez}}, \bibinfo {author}
  {\bibfnamefont {A.}~\bibnamefont {Varas}}, \bibinfo {author} {\bibfnamefont
  {I.}~\bibnamefont {Theophilou}}, \bibinfo {author} {\bibfnamefont
  {N.}~\bibnamefont {Helbig}}, \bibinfo {author} {\bibfnamefont {M.~J.}\
  \bibnamefont {Verstraete}}, \bibinfo {author} {\bibfnamefont
  {L.}~\bibnamefont {Stella}}, \bibinfo {author} {\bibfnamefont
  {F.}~\bibnamefont {Nogueira}}, \bibinfo {author} {\bibfnamefont
  {A.}~\bibnamefont {Aspuru-Guzik}}, \bibinfo {author} {\bibfnamefont
  {A.}~\bibnamefont {Castro}}, \bibinfo {author} {\bibfnamefont {M.~A.~L.}\
  \bibnamefont {Marques}},\ and\ \bibinfo {author} {\bibfnamefont
  {A.}~\bibnamefont {Rubio}},\ }\bibfield  {title} {\bibinfo {title}
  {Real-space grids and the octopus code as tools for the development of new
  simulation approaches for electronic systems},\ }\href
  {https://doi.org/10.1039/C5CP00351B} {\bibfield  {journal} {\bibinfo
  {journal} {Phys.~Chem.~Chem.~Phys.~}\ }\textbf {\bibinfo {volume} {17}},\
  \bibinfo {pages} {31371} (\bibinfo {year} {2015})}\BibitemShut {NoStop}%
\bibitem [{\citenamefont {Tancogne-Dejean}\ \emph {et~al.}(2020)\citenamefont
  {Tancogne-Dejean}, \citenamefont {Oliveira}, \citenamefont {Andrade},
  \citenamefont {Appel}, \citenamefont {Borca}, \citenamefont {Le~Breton},
  \citenamefont {Buchholz}, \citenamefont {Castro}, \citenamefont {Corni},
  \citenamefont {Correa}, \citenamefont {De~Giovannini}, \citenamefont
  {Delgado}, \citenamefont {Eich}, \citenamefont {Flick}, \citenamefont {Gil},
  \citenamefont {Gomez}, \citenamefont {Helbig}, \citenamefont {Hübener},
  \citenamefont {Jestädt}, \citenamefont {Jornet-Somoza}, \citenamefont
  {Larsen}, \citenamefont {Lebedeva}, \citenamefont {Lüders}, \citenamefont
  {Marques}, \citenamefont {Ohlmann}, \citenamefont {Pipolo}, \citenamefont
  {Rampp}, \citenamefont {Rozzi}, \citenamefont {Strubbe}, \citenamefont
  {Sato}, \citenamefont {Schäfer}, \citenamefont {Theophilou}, \citenamefont
  {Welden},\ and\ \citenamefont {Rubio}}]{octopus2020}%
  \BibitemOpen
  \bibfield  {author} {\bibinfo {author} {\bibfnamefont {N.}~\bibnamefont
  {Tancogne-Dejean}}, \bibinfo {author} {\bibfnamefont {M.~J.~T.}\ \bibnamefont
  {Oliveira}}, \bibinfo {author} {\bibfnamefont {X.}~\bibnamefont {Andrade}},
  \bibinfo {author} {\bibfnamefont {H.}~\bibnamefont {Appel}}, \bibinfo
  {author} {\bibfnamefont {C.~H.}\ \bibnamefont {Borca}}, \bibinfo {author}
  {\bibfnamefont {G.}~\bibnamefont {Le~Breton}}, \bibinfo {author}
  {\bibfnamefont {F.}~\bibnamefont {Buchholz}}, \bibinfo {author}
  {\bibfnamefont {A.}~\bibnamefont {Castro}}, \bibinfo {author} {\bibfnamefont
  {S.}~\bibnamefont {Corni}}, \bibinfo {author} {\bibfnamefont {A.~A.}\
  \bibnamefont {Correa}}, \bibinfo {author} {\bibfnamefont {U.}~\bibnamefont
  {De~Giovannini}}, \bibinfo {author} {\bibfnamefont {A.}~\bibnamefont
  {Delgado}}, \bibinfo {author} {\bibfnamefont {F.~G.}\ \bibnamefont {Eich}},
  \bibinfo {author} {\bibfnamefont {J.}~\bibnamefont {Flick}}, \bibinfo
  {author} {\bibfnamefont {G.}~\bibnamefont {Gil}}, \bibinfo {author}
  {\bibfnamefont {A.}~\bibnamefont {Gomez}}, \bibinfo {author} {\bibfnamefont
  {N.}~\bibnamefont {Helbig}}, \bibinfo {author} {\bibfnamefont
  {H.}~\bibnamefont {Hübener}}, \bibinfo {author} {\bibfnamefont
  {R.}~\bibnamefont {Jestädt}}, \bibinfo {author} {\bibfnamefont
  {J.}~\bibnamefont {Jornet-Somoza}}, \bibinfo {author} {\bibfnamefont {A.~H.}\
  \bibnamefont {Larsen}}, \bibinfo {author} {\bibfnamefont {I.~V.}\
  \bibnamefont {Lebedeva}}, \bibinfo {author} {\bibfnamefont {M.}~\bibnamefont
  {Lüders}}, \bibinfo {author} {\bibfnamefont {M.~A.~L.}\ \bibnamefont
  {Marques}}, \bibinfo {author} {\bibfnamefont {S.~T.}\ \bibnamefont
  {Ohlmann}}, \bibinfo {author} {\bibfnamefont {S.}~\bibnamefont {Pipolo}},
  \bibinfo {author} {\bibfnamefont {M.}~\bibnamefont {Rampp}}, \bibinfo
  {author} {\bibfnamefont {C.~A.}\ \bibnamefont {Rozzi}}, \bibinfo {author}
  {\bibfnamefont {D.~A.}\ \bibnamefont {Strubbe}}, \bibinfo {author}
  {\bibfnamefont {S.~A.}\ \bibnamefont {Sato}}, \bibinfo {author}
  {\bibfnamefont {C.}~\bibnamefont {Schäfer}}, \bibinfo {author}
  {\bibfnamefont {I.}~\bibnamefont {Theophilou}}, \bibinfo {author}
  {\bibfnamefont {A.}~\bibnamefont {Welden}},\ and\ \bibinfo {author}
  {\bibfnamefont {A.}~\bibnamefont {Rubio}},\ }\bibfield  {title} {\bibinfo
  {title} {Octopus, a computational framework for exploring light-driven
  phenomena and quantum dynamics in extended and finite systems},\ }\href
  {https://doi.org/10.1063/1.5142502} {\bibfield  {journal} {\bibinfo
  {journal} {J.~Chem.~Phys.~}\ }\textbf {\bibinfo {volume} {152}},\ \bibinfo
  {pages} {124119} (\bibinfo {year} {2020})},\ \Eprint
  {https://arxiv.org/abs/https://doi.org/10.1063/1.5142502}
  {https://doi.org/10.1063/1.5142502} \BibitemShut {NoStop}%
\bibitem [{\citenamefont {Bitzek}\ \emph {et~al.}(2006)\citenamefont {Bitzek},
  \citenamefont {Koskinen}, \citenamefont {G\"ahler}, \citenamefont {Moseler},\
  and\ \citenamefont {Gumbsch}}]{fire}%
  \BibitemOpen
  \bibfield  {author} {\bibinfo {author} {\bibfnamefont {E.}~\bibnamefont
  {Bitzek}}, \bibinfo {author} {\bibfnamefont {P.}~\bibnamefont {Koskinen}},
  \bibinfo {author} {\bibfnamefont {F.}~\bibnamefont {G\"ahler}}, \bibinfo
  {author} {\bibfnamefont {M.}~\bibnamefont {Moseler}},\ and\ \bibinfo {author}
  {\bibfnamefont {P.}~\bibnamefont {Gumbsch}},\ }\bibfield  {title} {\bibinfo
  {title} {Structural relaxation made simple},\ }\href
  {https://doi.org/10.1103/PhysRevLett.97.170201} {\bibfield  {journal}
  {\bibinfo  {journal} {Phys.~Rev.~Lett.~}\ }\textbf {\bibinfo {volume} {97}},\
  \bibinfo {pages} {170201} (\bibinfo {year} {2006})}\BibitemShut {NoStop}%
\bibitem [{\citenamefont {Castro}\ \emph {et~al.}(2004)\citenamefont {Castro},
  \citenamefont {Marques},\ and\ \citenamefont {Rubio}}]{castro+2004jcp}%
  \BibitemOpen
  \bibfield  {author} {\bibinfo {author} {\bibfnamefont {A.}~\bibnamefont
  {Castro}}, \bibinfo {author} {\bibfnamefont {M.~A.~L.}\ \bibnamefont
  {Marques}},\ and\ \bibinfo {author} {\bibfnamefont {A.}~\bibnamefont
  {Rubio}},\ }\bibfield  {title} {\bibinfo {title} {Propagators for the
  time-dependent kohn–sham equations},\ }\href
  {https://doi.org/10.1063/1.1774980} {\bibfield  {journal} {\bibinfo
  {journal} {J.~Chem.~Phys.~}\ }\textbf {\bibinfo {volume} {121}},\ \bibinfo
  {pages} {3425} (\bibinfo {year} {2004})},\ \Eprint
  {https://arxiv.org/abs/https://doi.org/10.1063/1.1774980}
  {https://doi.org/10.1063/1.1774980} \BibitemShut {NoStop}%
\bibitem [{\citenamefont {Perdew}\ and\ \citenamefont {Zunger}(1981)}]{pz}%
  \BibitemOpen
  \bibfield  {author} {\bibinfo {author} {\bibfnamefont {J.~P.}\ \bibnamefont
  {Perdew}}\ and\ \bibinfo {author} {\bibfnamefont {A.}~\bibnamefont
  {Zunger}},\ }\bibfield  {title} {\bibinfo {title} {Self-interaction
  correction to density-functional approximations for many-electron systems},\
  }\href@noop {} {\bibfield  {journal} {\bibinfo  {journal} {Phys.~Rev.~B}\
  }\textbf {\bibinfo {volume} {23}},\ \bibinfo {pages} {5048} (\bibinfo {year}
  {1981})}\BibitemShut {NoStop}%
\bibitem [{\citenamefont {Troullier}\ and\ \citenamefont
  {Martins}(1991)}]{trou-mart91prb}%
  \BibitemOpen
  \bibfield  {author} {\bibinfo {author} {\bibfnamefont {N.}~\bibnamefont
  {Troullier}}\ and\ \bibinfo {author} {\bibfnamefont {J.~L.}\ \bibnamefont
  {Martins}},\ }\bibfield  {title} {\bibinfo {title} {Efficient
  pseudopotentials for plane-wave calculations},\ }\href@noop {} {\bibfield
  {journal} {\bibinfo  {journal} {Phys.~Rev.~B}\ }\textbf {\bibinfo {volume}
  {43}},\ \bibinfo {pages} {1993} (\bibinfo {year} {1991})}\BibitemShut
  {NoStop}%
\bibitem [{\citenamefont {Krumland}\ \emph {et~al.}(2021)\citenamefont
  {Krumland}, \citenamefont {Gil}, \citenamefont {Corni},\ and\ \citenamefont
  {Cocchi}}]{krum+21jcp}%
  \BibitemOpen
  \bibfield  {author} {\bibinfo {author} {\bibfnamefont {J.}~\bibnamefont
  {Krumland}}, \bibinfo {author} {\bibfnamefont {G.}~\bibnamefont {Gil}},
  \bibinfo {author} {\bibfnamefont {S.}~\bibnamefont {Corni}},\ and\ \bibinfo
  {author} {\bibfnamefont {C.}~\bibnamefont {Cocchi}},\ }\bibfield  {title}
  {\bibinfo {title} {Layerpcm: An implicit scheme for dielectric screening from
  layered substrates},\ }\href@noop {} {\bibfield  {journal} {\bibinfo
  {journal} {J.~Chem.~Phys.~}\ }\textbf {\bibinfo {volume} {154}},\ \bibinfo
  {pages} {224114} (\bibinfo {year} {2021})}\BibitemShut {NoStop}%
\bibitem [{\citenamefont {Koch}\ and\ \citenamefont
  {Otto}(1972)}]{kochOtto1972cpl}%
  \BibitemOpen
  \bibfield  {author} {\bibinfo {author} {\bibfnamefont {E.}~\bibnamefont
  {Koch}}\ and\ \bibinfo {author} {\bibfnamefont {A.}~\bibnamefont {Otto}},\
  }\bibfield  {title} {\bibinfo {title} {Optical absorption of benzene vapour
  for photon energies from 6 ev to 35 ev},\ }\href
  {https://doi.org/https://doi.org/10.1016/0009-2614(72)90011-5} {\bibfield
  {journal} {\bibinfo  {journal} {Chem.~Phys.~Lett.~}\ }\textbf {\bibinfo
  {volume} {12}},\ \bibinfo {pages} {476} (\bibinfo {year} {1972})}\BibitemShut
  {NoStop}%
\bibitem [{\citenamefont {Yabana}\ and\ \citenamefont
  {Bertsch}(1999)}]{yabanaBertsch1999ijqp}%
  \BibitemOpen
  \bibfield  {author} {\bibinfo {author} {\bibfnamefont {K.}~\bibnamefont
  {Yabana}}\ and\ \bibinfo {author} {\bibfnamefont {G.~F.}\ \bibnamefont
  {Bertsch}},\ }\bibfield  {title} {\bibinfo {title} {Time-dependent
  local-density approximation in real time: Application to conjugated
  molecules},\ }\href {https://www.osti.gov/biblio/687545} {\bibfield
  {journal} {\bibinfo  {journal} {Int.~J.~Quantum~Chem.}\ }\textbf {\bibinfo
  {volume} {75}} (\bibinfo {year} {1999})}\BibitemShut {NoStop}%
\bibitem [{\citenamefont {Harris}\ and\ \citenamefont
  {Bertolucci}(1989)}]{harris1989}%
  \BibitemOpen
  \bibfield  {author} {\bibinfo {author} {\bibfnamefont {D.~C.}\ \bibnamefont
  {Harris}}\ and\ \bibinfo {author} {\bibfnamefont {M.~D.}\ \bibnamefont
  {Bertolucci}},\ }\href@noop {} {\emph {\bibinfo {title} {Symmetry and
  spectroscopy: an introduction to vibrational and electronic spectroscopy}}}\
  (\bibinfo  {publisher} {Courier Corporation},\ \bibinfo {year}
  {1989})\BibitemShut {NoStop}%
\bibitem [{\citenamefont {Cocchi}\ \emph
  {et~al.}(2014{\natexlab{a}})\citenamefont {Cocchi}, \citenamefont {Prezzi},
  \citenamefont {Ruini}, \citenamefont {Caldas},\ and\ \citenamefont
  {Molinari}}]{cocchi+2014jpca}%
  \BibitemOpen
  \bibfield  {author} {\bibinfo {author} {\bibfnamefont {C.}~\bibnamefont
  {Cocchi}}, \bibinfo {author} {\bibfnamefont {D.}~\bibnamefont {Prezzi}},
  \bibinfo {author} {\bibfnamefont {A.}~\bibnamefont {Ruini}}, \bibinfo
  {author} {\bibfnamefont {M.~J.}\ \bibnamefont {Caldas}},\ and\ \bibinfo
  {author} {\bibfnamefont {E.}~\bibnamefont {Molinari}},\ }\bibfield  {title}
  {\bibinfo {title} {Anisotropy and size effects on the optical spectra of
  polycyclic aromatic hydrocarbons},\ }\href
  {https://doi.org/10.1021/jp503054j} {\bibfield  {journal} {\bibinfo
  {journal} {J.~Phys.~Chem.~A}\ }\textbf {\bibinfo {volume} {118}},\ \bibinfo
  {pages} {6507} (\bibinfo {year} {2014}{\natexlab{a}})},\ \bibinfo {note}
  {pMID: 24984100},\ \Eprint
  {https://arxiv.org/abs/https://doi.org/10.1021/jp503054j}
  {https://doi.org/10.1021/jp503054j} \BibitemShut {NoStop}%
\bibitem [{\citenamefont {Wilson}(1934)}]{wilson1934pr}%
  \BibitemOpen
  \bibfield  {author} {\bibinfo {author} {\bibfnamefont {E.~B.}\ \bibnamefont
  {Wilson}},\ }\bibfield  {title} {\bibinfo {title} {The normal modes and
  frequencies of vibration of the regular plane hexagon model of the benzene
  molecule},\ }\href {https://doi.org/10.1103/PhysRev.45.706} {\bibfield
  {journal} {\bibinfo  {journal} {Phys.~Rev.~}\ }\textbf {\bibinfo {volume}
  {45}},\ \bibinfo {pages} {706} (\bibinfo {year} {1934})}\BibitemShut
  {NoStop}%
\bibitem [{\citenamefont {Barbatti}\ \emph {et~al.}(2010)\citenamefont
  {Barbatti}, \citenamefont {Aquino},\ and\ \citenamefont
  {Lischka}}]{barbatti+2010pccp}%
  \BibitemOpen
  \bibfield  {author} {\bibinfo {author} {\bibfnamefont {M.}~\bibnamefont
  {Barbatti}}, \bibinfo {author} {\bibfnamefont {A.~J.~A.}\ \bibnamefont
  {Aquino}},\ and\ \bibinfo {author} {\bibfnamefont {H.}~\bibnamefont
  {Lischka}},\ }\bibfield  {title} {\bibinfo {title} {The uv absorption of
  nucleobases: semi-classical ab initio spectra simulations},\ }\href
  {https://doi.org/10.1039/B924956G} {\bibfield  {journal} {\bibinfo  {journal}
  {Phys.~Chem.~Chem.~Phys.~}\ }\textbf {\bibinfo {volume} {12}},\ \bibinfo
  {pages} {4959} (\bibinfo {year} {2010})}\BibitemShut {NoStop}%
\bibitem [{\citenamefont {Crespo-Otero}\ and\ \citenamefont
  {Barbatti}(2012)}]{crespo-oteroBarbatti2012tcacc}%
  \BibitemOpen
  \bibfield  {author} {\bibinfo {author} {\bibfnamefont {R.}~\bibnamefont
  {Crespo-Otero}}\ and\ \bibinfo {author} {\bibfnamefont {M.}~\bibnamefont
  {Barbatti}},\ }\bibfield  {title} {\bibinfo {title} {Spectrum simulation and
  decomposition with nuclear ensemble: formal derivation and application to
  benzene, furan and 2-phenylfuran},\ }\href
  {https://doi.org/10.1007/s00214-012-1237-4} {\bibfield  {journal} {\bibinfo
  {journal} {Theo.~Chem.~Accounts}\ }\textbf {\bibinfo {volume} {131}},\
  \bibinfo {pages} {1237} (\bibinfo {year} {2012})}\BibitemShut {NoStop}%
\bibitem [{\citenamefont {Lively}\ \emph {et~al.}(2021)\citenamefont {Lively},
  \citenamefont {Albareda}, \citenamefont {Sato}, \citenamefont {Kelly},\ and\
  \citenamefont {Rubio}}]{lively+2021jpcl}%
  \BibitemOpen
  \bibfield  {author} {\bibinfo {author} {\bibfnamefont {K.}~\bibnamefont
  {Lively}}, \bibinfo {author} {\bibfnamefont {G.}~\bibnamefont {Albareda}},
  \bibinfo {author} {\bibfnamefont {S.~A.}\ \bibnamefont {Sato}}, \bibinfo
  {author} {\bibfnamefont {A.}~\bibnamefont {Kelly}},\ and\ \bibinfo {author}
  {\bibfnamefont {A.}~\bibnamefont {Rubio}},\ }\bibfield  {title} {\bibinfo
  {title} {Simulating vibronic spectra without born–oppenheimer surfaces},\
  }\href {https://doi.org/10.1021/acs.jpclett.1c00073} {\bibfield  {journal}
  {\bibinfo  {journal} {J.~Phys.~Chem.~Lett.}\ }\textbf {\bibinfo {volume}
  {12}},\ \bibinfo {pages} {3074} (\bibinfo {year} {2021})},\ \bibinfo {note}
  {pMID: 33750137},\ \Eprint
  {https://arxiv.org/abs/https://doi.org/10.1021/acs.jpclett.1c00073}
  {https://doi.org/10.1021/acs.jpclett.1c00073} \BibitemShut {NoStop}%
\bibitem [{\citenamefont {Hollas}(1964)}]{clar1964pah}%
  \BibitemOpen
  \bibfield  {author} {\bibinfo {author} {\bibfnamefont {J.~M.}\ \bibnamefont
  {Hollas}},\ }\href@noop {} {\emph {\bibinfo {title} {Polycyclic
  Hydrocarbons}}}\ (\bibinfo  {publisher} {Springer, Berlin},\ \bibinfo {year}
  {1964})\BibitemShut {NoStop}%
\bibitem [{\citenamefont {Ohno}\ \emph {et~al.}(1972)\citenamefont {Ohno},
  \citenamefont {Kajiwara},\ and\ \citenamefont {Inokuchi}}]{ohno+1972bull}%
  \BibitemOpen
  \bibfield  {author} {\bibinfo {author} {\bibfnamefont {K.}~\bibnamefont
  {Ohno}}, \bibinfo {author} {\bibfnamefont {T.}~\bibnamefont {Kajiwara}},\
  and\ \bibinfo {author} {\bibfnamefont {H.}~\bibnamefont {Inokuchi}},\
  }\bibfield  {title} {\bibinfo {title} {Vibrational analysis of electronic
  transition bands of coronene},\ }\href {https://doi.org/10.1246/bcsj.45.996}
  {\bibfield  {journal} {\bibinfo  {journal} {Bull.~Chem.~Soc.~Jpn.}\ }\textbf
  {\bibinfo {volume} {45}},\ \bibinfo {pages} {996} (\bibinfo {year} {1972})},\
  \Eprint {https://arxiv.org/abs/https://doi.org/10.1246/bcsj.45.996}
  {https://doi.org/10.1246/bcsj.45.996} \BibitemShut {NoStop}%
\bibitem [{\citenamefont {Cataldo}\ \emph {et~al.}(2011)\citenamefont
  {Cataldo}, \citenamefont {Ursini}, \citenamefont {Angelini},\ and\
  \citenamefont {Iglesias-Groth}}]{cataldo+2011full}%
  \BibitemOpen
  \bibfield  {author} {\bibinfo {author} {\bibfnamefont {F.}~\bibnamefont
  {Cataldo}}, \bibinfo {author} {\bibfnamefont {O.}~\bibnamefont {Ursini}},
  \bibinfo {author} {\bibfnamefont {G.}~\bibnamefont {Angelini}},\ and\
  \bibinfo {author} {\bibfnamefont {S.}~\bibnamefont {Iglesias-Groth}},\
  }\bibfield  {title} {\bibinfo {title} {On the way to graphene: The bottom-up
  approach to very large pahs using the scholl reaction},\ }\href
  {https://doi.org/10.1080/1536383X.2010.494787} {\bibfield  {journal}
  {\bibinfo  {journal} {Fuller.~Nanotub}\ }\textbf {\bibinfo {volume} {19}},\
  \bibinfo {pages} {713} (\bibinfo {year} {2011})},\ \Eprint
  {https://arxiv.org/abs/https://doi.org/10.1080/1536383X.2010.494787}
  {https://doi.org/10.1080/1536383X.2010.494787} \BibitemShut {NoStop}%
\bibitem [{\citenamefont {Min}\ \emph {et~al.}(2017)\citenamefont {Min},
  \citenamefont {Agostini}, \citenamefont {Tavernelli},\ and\ \citenamefont
  {Gross}}]{min+2017jpcl}%
  \BibitemOpen
  \bibfield  {author} {\bibinfo {author} {\bibfnamefont {S.~K.}\ \bibnamefont
  {Min}}, \bibinfo {author} {\bibfnamefont {F.}~\bibnamefont {Agostini}},
  \bibinfo {author} {\bibfnamefont {I.}~\bibnamefont {Tavernelli}},\ and\
  \bibinfo {author} {\bibfnamefont {E.~K.~U.}\ \bibnamefont {Gross}},\
  }\bibfield  {title} {\bibinfo {title} {Ab initio nonadiabatic dynamics with
  coupled trajectories: A rigorous approach to quantum (de)coherence},\ }\href
  {https://doi.org/10.1021/acs.jpclett.7b01249} {\bibfield  {journal} {\bibinfo
   {journal} {J.~Phys.~Chem.~Lett.}\ }\textbf {\bibinfo {volume} {8}},\
  \bibinfo {pages} {3048} (\bibinfo {year} {2017})},\ \bibinfo {note} {pMID:
  28618782},\ \Eprint
  {https://arxiv.org/abs/https://doi.org/10.1021/acs.jpclett.7b01249}
  {https://doi.org/10.1021/acs.jpclett.7b01249} \BibitemShut {NoStop}%
\bibitem [{\citenamefont {Albareda}\ \emph {et~al.}(2021)\citenamefont
  {Albareda}, \citenamefont {Lively}, \citenamefont {Sato}, \citenamefont
  {Kelly},\ and\ \citenamefont {Rubio}}]{albareda2021jctc}%
  \BibitemOpen
  \bibfield  {author} {\bibinfo {author} {\bibfnamefont {G.}~\bibnamefont
  {Albareda}}, \bibinfo {author} {\bibfnamefont {K.}~\bibnamefont {Lively}},
  \bibinfo {author} {\bibfnamefont {S.~A.}\ \bibnamefont {Sato}}, \bibinfo
  {author} {\bibfnamefont {A.}~\bibnamefont {Kelly}},\ and\ \bibinfo {author}
  {\bibfnamefont {A.}~\bibnamefont {Rubio}},\ }\bibfield  {title} {\bibinfo
  {title} {Conditional wave function theory: A unified treatment of molecular
  structure and nonadiabatic dynamics},\ }\href@noop {} {\bibfield  {journal}
  {\bibinfo  {journal} {J.~Chem.~Theory.~Comput.~}\ }\textbf {\bibinfo {volume}
  {17}},\ \bibinfo {pages} {7321} (\bibinfo {year} {2021})}\BibitemShut
  {NoStop}%
\bibitem [{\citenamefont {Heller}(1976{\natexlab{b}})}]{heller1976jcp}%
  \BibitemOpen
  \bibfield  {author} {\bibinfo {author} {\bibfnamefont {E.~J.}\ \bibnamefont
  {Heller}},\ }\bibfield  {title} {\bibinfo {title} {Wigner phase space method:
  Analysis for semiclassical applications},\ }\href
  {https://doi.org/10.1063/1.433238} {\bibfield  {journal} {\bibinfo  {journal}
  {J.~Chem.~Phys.~}\ }\textbf {\bibinfo {volume} {65}},\ \bibinfo {pages}
  {1289} (\bibinfo {year} {1976}{\natexlab{b}})},\ \Eprint
  {https://arxiv.org/abs/https://doi.org/10.1063/1.433238}
  {https://doi.org/10.1063/1.433238} \BibitemShut {NoStop}%
\bibitem [{\citenamefont {Kuda-Singappulige}\ \emph {et~al.}(2020)\citenamefont
  {Kuda-Singappulige}, \citenamefont {Wildman}, \citenamefont {Lingerfelt},
  \citenamefont {Li},\ and\ \citenamefont {Aikens}}]{kuda+2020jpca}%
  \BibitemOpen
  \bibfield  {author} {\bibinfo {author} {\bibfnamefont {G.~U.}\ \bibnamefont
  {Kuda-Singappulige}}, \bibinfo {author} {\bibfnamefont {A.}~\bibnamefont
  {Wildman}}, \bibinfo {author} {\bibfnamefont {D.~B.}\ \bibnamefont
  {Lingerfelt}}, \bibinfo {author} {\bibfnamefont {X.}~\bibnamefont {Li}},\
  and\ \bibinfo {author} {\bibfnamefont {C.~M.}\ \bibnamefont {Aikens}},\
  }\bibfield  {title} {\bibinfo {title} {Ultrafast nonradiative decay of a
  dipolar plasmon-like state in naphthalene},\ }\href
  {https://doi.org/10.1021/acs.jpca.0c09564} {\bibfield  {journal} {\bibinfo
  {journal} {J.~Phys.~Chem.~A}\ }\textbf {\bibinfo {volume} {124}},\ \bibinfo
  {pages} {9729} (\bibinfo {year} {2020})},\ \bibinfo {note} {pMID: 33181013},\
  \Eprint {https://arxiv.org/abs/https://doi.org/10.1021/acs.jpca.0c09564}
  {https://doi.org/10.1021/acs.jpca.0c09564} \BibitemShut {NoStop}%
\bibitem [{\citenamefont {Brixner}\ \emph {et~al.}(2005)\citenamefont
  {Brixner}, \citenamefont {Stenger}, \citenamefont {Vaswani}, \citenamefont
  {Cho}, \citenamefont {Blankenship},\ and\ \citenamefont
  {Fleming}}]{brixner+2005nature}%
  \BibitemOpen
  \bibfield  {author} {\bibinfo {author} {\bibfnamefont {T.}~\bibnamefont
  {Brixner}}, \bibinfo {author} {\bibfnamefont {J.}~\bibnamefont {Stenger}},
  \bibinfo {author} {\bibfnamefont {H.~M.}\ \bibnamefont {Vaswani}}, \bibinfo
  {author} {\bibfnamefont {M.}~\bibnamefont {Cho}}, \bibinfo {author}
  {\bibfnamefont {R.~E.}\ \bibnamefont {Blankenship}},\ and\ \bibinfo {author}
  {\bibfnamefont {G.~R.}\ \bibnamefont {Fleming}},\ }\bibfield  {title}
  {\bibinfo {title} {Two-dimensional spectroscopy of electronic couplings in
  photosynthesis},\ }\href {https://doi.org/10.1038/nature03429} {\bibfield
  {journal} {\bibinfo  {journal} {Nature~(London)~}\ }\textbf {\bibinfo
  {volume} {434}},\ \bibinfo {pages} {625} (\bibinfo {year}
  {2005})}\BibitemShut {NoStop}%
\bibitem [{\citenamefont {Engel}\ \emph {et~al.}(2007)\citenamefont {Engel},
  \citenamefont {Calhoun}, \citenamefont {Read}, \citenamefont {Ahn},
  \citenamefont {Man{\v{c}}al}, \citenamefont {Cheng}, \citenamefont
  {Blankenship},\ and\ \citenamefont {Fleming}}]{engel+2007nature}%
  \BibitemOpen
  \bibfield  {author} {\bibinfo {author} {\bibfnamefont {G.~S.}\ \bibnamefont
  {Engel}}, \bibinfo {author} {\bibfnamefont {T.~R.}\ \bibnamefont {Calhoun}},
  \bibinfo {author} {\bibfnamefont {E.~L.}\ \bibnamefont {Read}}, \bibinfo
  {author} {\bibfnamefont {T.-K.}\ \bibnamefont {Ahn}}, \bibinfo {author}
  {\bibfnamefont {T.}~\bibnamefont {Man{\v{c}}al}}, \bibinfo {author}
  {\bibfnamefont {Y.-C.}\ \bibnamefont {Cheng}}, \bibinfo {author}
  {\bibfnamefont {R.~E.}\ \bibnamefont {Blankenship}},\ and\ \bibinfo {author}
  {\bibfnamefont {G.~R.}\ \bibnamefont {Fleming}},\ }\bibfield  {title}
  {\bibinfo {title} {Evidence for wavelike energy transfer through quantum
  coherence in photosynthetic systems},\ }\href
  {https://doi.org/10.1038/nature05678} {\bibfield  {journal} {\bibinfo
  {journal} {Nature~(London)~}\ }\textbf {\bibinfo {volume} {446}},\ \bibinfo
  {pages} {782} (\bibinfo {year} {2007})}\BibitemShut {NoStop}%
\bibitem [{\citenamefont {Lee}\ \emph {et~al.}(2007)\citenamefont {Lee},
  \citenamefont {Cheng},\ and\ \citenamefont {Fleming}}]{lee+2007science}%
  \BibitemOpen
  \bibfield  {author} {\bibinfo {author} {\bibfnamefont {H.}~\bibnamefont
  {Lee}}, \bibinfo {author} {\bibfnamefont {Y.~C.}\ \bibnamefont {Cheng}},\
  and\ \bibinfo {author} {\bibfnamefont {G.~R.}\ \bibnamefont {Fleming}},\
  }\bibfield  {title} {\bibinfo {title} {{{C}oherence dynamics in
  photosynthesis: protein protection of excitonic coherence}},\ }\href@noop {}
  {\bibfield  {journal} {\bibinfo  {journal} {Science}\ }\textbf {\bibinfo
  {volume} {316}},\ \bibinfo {pages} {1462} (\bibinfo {year}
  {2007})}\BibitemShut {NoStop}%
\bibitem [{\citenamefont {Collini}\ \emph {et~al.}(2010)\citenamefont
  {Collini}, \citenamefont {Wong}, \citenamefont {Wilk}, \citenamefont {Curmi},
  \citenamefont {Brumer},\ and\ \citenamefont {Scholes}}]{collini+2010nature}%
  \BibitemOpen
  \bibfield  {author} {\bibinfo {author} {\bibfnamefont {E.}~\bibnamefont
  {Collini}}, \bibinfo {author} {\bibfnamefont {C.~Y.}\ \bibnamefont {Wong}},
  \bibinfo {author} {\bibfnamefont {K.~E.}\ \bibnamefont {Wilk}}, \bibinfo
  {author} {\bibfnamefont {P.~M.~G.}\ \bibnamefont {Curmi}}, \bibinfo {author}
  {\bibfnamefont {P.}~\bibnamefont {Brumer}},\ and\ \bibinfo {author}
  {\bibfnamefont {G.~D.}\ \bibnamefont {Scholes}},\ }\bibfield  {title}
  {\bibinfo {title} {Coherently wired light-harvesting in photosynthetic marine
  algae at ambient temperature},\ }\href {https://doi.org/10.1038/nature08811}
  {\bibfield  {journal} {\bibinfo  {journal} {Nature~(London)~}\ }\textbf
  {\bibinfo {volume} {463}},\ \bibinfo {pages} {644} (\bibinfo {year}
  {2010})}\BibitemShut {NoStop}%
\bibitem [{\citenamefont {Panitchayangkoon}\ \emph {et~al.}(2010)\citenamefont
  {Panitchayangkoon}, \citenamefont {Hayes}, \citenamefont {Fransted},
  \citenamefont {Caram}, \citenamefont {Harel}, \citenamefont {Wen},
  \citenamefont {Blankenship},\ and\ \citenamefont
  {Engel}}]{pani+2010proceedings}%
  \BibitemOpen
  \bibfield  {author} {\bibinfo {author} {\bibfnamefont {G.}~\bibnamefont
  {Panitchayangkoon}}, \bibinfo {author} {\bibfnamefont {D.}~\bibnamefont
  {Hayes}}, \bibinfo {author} {\bibfnamefont {K.~A.}\ \bibnamefont {Fransted}},
  \bibinfo {author} {\bibfnamefont {J.~R.}\ \bibnamefont {Caram}}, \bibinfo
  {author} {\bibfnamefont {E.}~\bibnamefont {Harel}}, \bibinfo {author}
  {\bibfnamefont {J.}~\bibnamefont {Wen}}, \bibinfo {author} {\bibfnamefont
  {R.~E.}\ \bibnamefont {Blankenship}},\ and\ \bibinfo {author} {\bibfnamefont
  {G.~S.}\ \bibnamefont {Engel}},\ }\bibfield  {title} {\bibinfo {title}
  {Long-lived quantum coherence in photosynthetic complexes at physiological
  temperature},\ }\href@noop {} {\bibfield  {journal} {\bibinfo  {journal}
  {Proc. Natl. Acad. Sci. USA}\ }\textbf {\bibinfo {volume} {107}},\ \bibinfo
  {pages} {12766} (\bibinfo {year} {2010})}\BibitemShut {NoStop}%
\bibitem [{\citenamefont {Hildner}\ \emph {et~al.}(2013)\citenamefont
  {Hildner}, \citenamefont {Brinks}, \citenamefont {Nieder}, \citenamefont
  {Cogdell},\ and\ \citenamefont {van Hulst}}]{hildner+2013sci}%
  \BibitemOpen
  \bibfield  {author} {\bibinfo {author} {\bibfnamefont {R.}~\bibnamefont
  {Hildner}}, \bibinfo {author} {\bibfnamefont {D.}~\bibnamefont {Brinks}},
  \bibinfo {author} {\bibfnamefont {J.~B.}\ \bibnamefont {Nieder}}, \bibinfo
  {author} {\bibfnamefont {R.~J.}\ \bibnamefont {Cogdell}},\ and\ \bibinfo
  {author} {\bibfnamefont {N.~F.}\ \bibnamefont {van Hulst}},\ }\bibfield
  {title} {\bibinfo {title} {{{Q}uantum coherent energy transfer over varying
  pathways in single light-harvesting complexes}},\ }\href@noop {} {\bibfield
  {journal} {\bibinfo  {journal} {Science}\ }\textbf {\bibinfo {volume}
  {340}},\ \bibinfo {pages} {1448} (\bibinfo {year} {2013})}\BibitemShut
  {NoStop}%
\bibitem [{\citenamefont {Seidner}\ \emph {et~al.}(1995)\citenamefont
  {Seidner}, \citenamefont {Stock},\ and\ \citenamefont
  {Domcke}}]{seidner+1995jcp}%
  \BibitemOpen
  \bibfield  {author} {\bibinfo {author} {\bibfnamefont {L.}~\bibnamefont
  {Seidner}}, \bibinfo {author} {\bibfnamefont {G.}~\bibnamefont {Stock}},\
  and\ \bibinfo {author} {\bibfnamefont {W.}~\bibnamefont {Domcke}},\
  }\bibfield  {title} {\bibinfo {title} {Nonperturbative approach to
  femtosecond spectroscopy: General theory and application to multidimensional
  nonadiabatic photoisomerization processes},\ }\href
  {https://doi.org/10.1063/1.469586} {\bibfield  {journal} {\bibinfo  {journal}
  {J.~Chem.~Phys.~}\ }\textbf {\bibinfo {volume} {103}},\ \bibinfo {pages}
  {3998} (\bibinfo {year} {1995})},\ \Eprint
  {https://arxiv.org/abs/https://doi.org/10.1063/1.469586}
  {https://doi.org/10.1063/1.469586} \BibitemShut {NoStop}%
\bibitem [{\citenamefont {De~Giovannini}\ \emph {et~al.}(2013)\citenamefont
  {De~Giovannini}, \citenamefont {Brunetto}, \citenamefont {Castro},
  \citenamefont {Walkenhorst},\ and\ \citenamefont {Rubio}}]{umberto+2013cpc}%
  \BibitemOpen
  \bibfield  {author} {\bibinfo {author} {\bibfnamefont {U.}~\bibnamefont
  {De~Giovannini}}, \bibinfo {author} {\bibfnamefont {G.}~\bibnamefont
  {Brunetto}}, \bibinfo {author} {\bibfnamefont {A.}~\bibnamefont {Castro}},
  \bibinfo {author} {\bibfnamefont {J.}~\bibnamefont {Walkenhorst}},\ and\
  \bibinfo {author} {\bibfnamefont {A.}~\bibnamefont {Rubio}},\ }\bibfield
  {title} {\bibinfo {title} {Simulating pump–probe photoelectron and
  absorption spectroscopy on the attosecond timescale with time-dependent
  density functional theory},\ }\href
  {https://doi.org/https://doi.org/10.1002/cphc.201201007} {\bibfield
  {journal} {\bibinfo  {journal} {Comput.~Phys.~Commun.~}\ }\textbf {\bibinfo
  {volume} {14}},\ \bibinfo {pages} {1363} (\bibinfo {year}
  {2013})}\BibitemShut {NoStop}%
\bibitem [{\citenamefont {Bonafé}\ \emph {et~al.}(2018)\citenamefont
  {Bonafé}, \citenamefont {Hernández}, \citenamefont {Aradi}, \citenamefont
  {Frauenheim},\ and\ \citenamefont {Sánchez}}]{bonafe+2018jpcl}%
  \BibitemOpen
  \bibfield  {author} {\bibinfo {author} {\bibfnamefont {F.~P.}\ \bibnamefont
  {Bonafé}}, \bibinfo {author} {\bibfnamefont {F.~J.}\ \bibnamefont
  {Hernández}}, \bibinfo {author} {\bibfnamefont {B.}~\bibnamefont {Aradi}},
  \bibinfo {author} {\bibfnamefont {T.}~\bibnamefont {Frauenheim}},\ and\
  \bibinfo {author} {\bibfnamefont {C.~G.}\ \bibnamefont {Sánchez}},\
  }\bibfield  {title} {\bibinfo {title} {Fully atomistic real-time simulations
  of transient absorption spectroscopy},\ }\href
  {https://doi.org/10.1021/acs.jpclett.8b01659} {\bibfield  {journal} {\bibinfo
   {journal} {J.~Phys.~Chem.~Lett.}\ }\textbf {\bibinfo {volume} {9}},\
  \bibinfo {pages} {4355} (\bibinfo {year} {2018})},\ \bibinfo {note} {pMID:
  30024765},\ \Eprint
  {https://arxiv.org/abs/https://doi.org/10.1021/acs.jpclett.8b01659}
  {https://doi.org/10.1021/acs.jpclett.8b01659} \BibitemShut {NoStop}%
\bibitem [{\citenamefont {Hernández}\ \emph {et~al.}(2019)\citenamefont
  {Hernández}, \citenamefont {Bonafé}, \citenamefont {Aradi}, \citenamefont
  {Frauenheim},\ and\ \citenamefont {Sánchez}}]{hernandez+2019jpca}%
  \BibitemOpen
  \bibfield  {author} {\bibinfo {author} {\bibfnamefont {F.~J.}\ \bibnamefont
  {Hernández}}, \bibinfo {author} {\bibfnamefont {F.~P.}\ \bibnamefont
  {Bonafé}}, \bibinfo {author} {\bibfnamefont {B.}~\bibnamefont {Aradi}},
  \bibinfo {author} {\bibfnamefont {T.}~\bibnamefont {Frauenheim}},\ and\
  \bibinfo {author} {\bibfnamefont {C.~G.}\ \bibnamefont {Sánchez}},\
  }\bibfield  {title} {\bibinfo {title} {Simulation of impulsive vibrational
  spectroscopy},\ }\href {https://doi.org/10.1021/acs.jpca.9b00307} {\bibfield
  {journal} {\bibinfo  {journal} {J.~Phys.~Chem.~A}\ }\textbf {\bibinfo
  {volume} {123}},\ \bibinfo {pages} {2065} (\bibinfo {year} {2019})},\
  \bibinfo {note} {pMID: 30767532},\ \Eprint
  {https://arxiv.org/abs/https://doi.org/10.1021/acs.jpca.9b00307}
  {https://doi.org/10.1021/acs.jpca.9b00307} \BibitemShut {NoStop}%
\bibitem [{\citenamefont {Krumland}\ \emph {et~al.}(2020)\citenamefont
  {Krumland}, \citenamefont {Valencia}, \citenamefont {Pittalis}, \citenamefont
  {Rozzi},\ and\ \citenamefont {Cocchi}}]{krumland+2020jcp}%
  \BibitemOpen
  \bibfield  {author} {\bibinfo {author} {\bibfnamefont {J.}~\bibnamefont
  {Krumland}}, \bibinfo {author} {\bibfnamefont {A.~M.}\ \bibnamefont
  {Valencia}}, \bibinfo {author} {\bibfnamefont {S.}~\bibnamefont {Pittalis}},
  \bibinfo {author} {\bibfnamefont {C.~A.}\ \bibnamefont {Rozzi}},\ and\
  \bibinfo {author} {\bibfnamefont {C.}~\bibnamefont {Cocchi}},\ }\bibfield
  {title} {\bibinfo {title} {Understanding real-time time-dependent
  density-functional theory simulations of ultrafast laser-induced dynamics in
  organic molecules},\ }\href {https://doi.org/10.1063/5.0008194} {\bibfield
  {journal} {\bibinfo  {journal} {J.~Chem.~Phys.~}\ }\textbf {\bibinfo {volume}
  {153}},\ \bibinfo {pages} {054106} (\bibinfo {year} {2020})},\ \Eprint
  {https://arxiv.org/abs/https://doi.org/10.1063/5.0008194}
  {https://doi.org/10.1063/5.0008194} \BibitemShut {NoStop}%
\bibitem [{\citenamefont {Herperger}\ \emph {et~al.}(2021)\citenamefont
  {Herperger}, \citenamefont {Krumland},\ and\ \citenamefont
  {Cocchi}}]{herperger+2021jpca}%
  \BibitemOpen
  \bibfield  {author} {\bibinfo {author} {\bibfnamefont {K.~R.}\ \bibnamefont
  {Herperger}}, \bibinfo {author} {\bibfnamefont {J.}~\bibnamefont
  {Krumland}},\ and\ \bibinfo {author} {\bibfnamefont {C.}~\bibnamefont
  {Cocchi}},\ }\bibfield  {title} {\bibinfo {title} {Laser-induced electronic
  and vibronic dynamics in the pyrene molecule and its cation},\ }\href
  {https://doi.org/10.1021/acs.jpca.1c06538} {\bibfield  {journal} {\bibinfo
  {journal} {J.~Phys.~Chem.~A}\ }\textbf {\bibinfo {volume} {125}},\ \bibinfo
  {pages} {9619} (\bibinfo {year} {2021})},\ \bibinfo {note} {pMID: 34714646},\
  \Eprint {https://arxiv.org/abs/https://doi.org/10.1021/acs.jpca.1c06538}
  {https://doi.org/10.1021/acs.jpca.1c06538} \BibitemShut {NoStop}%
\bibitem [{\citenamefont {Schleich}(2011)}]{schleich2011quantum}%
  \BibitemOpen
  \bibfield  {author} {\bibinfo {author} {\bibfnamefont {W.}~\bibnamefont
  {Schleich}},\ }\href {https://books.google.de/books?id=2jUjQPW-WXAC} {\emph
  {\bibinfo {title} {Quantum Optics in Phase Space}}}\ (\bibinfo  {publisher}
  {Wiley},\ \bibinfo {year} {2011})\BibitemShut {NoStop}%
\bibitem [{\citenamefont {Miller}(2012)}]{miller2012jcp}%
  \BibitemOpen
  \bibfield  {author} {\bibinfo {author} {\bibfnamefont {W.~H.}\ \bibnamefont
  {Miller}},\ }\bibfield  {title} {\bibinfo {title} {Perspective: Quantum or
  classical coherence?},\ }\href {https://doi.org/10.1063/1.4727849} {\bibfield
   {journal} {\bibinfo  {journal} {J.~Chem.~Phys.~}\ }\textbf {\bibinfo
  {volume} {136}},\ \bibinfo {pages} {210901} (\bibinfo {year} {2012})},\
  \Eprint {https://arxiv.org/abs/https://doi.org/10.1063/1.4727849}
  {https://doi.org/10.1063/1.4727849} \BibitemShut {NoStop}%
\bibitem [{\citenamefont {Li}\ \emph {et~al.}(2010)\citenamefont {Li},
  \citenamefont {Lin}, \citenamefont {Li}, \citenamefont {Zhu},\ and\
  \citenamefont {Lin}}]{li+2010pccp}%
  \BibitemOpen
  \bibfield  {author} {\bibinfo {author} {\bibfnamefont {J.}~\bibnamefont
  {Li}}, \bibinfo {author} {\bibfnamefont {C.-K.}\ \bibnamefont {Lin}},
  \bibinfo {author} {\bibfnamefont {X.~Y.}\ \bibnamefont {Li}}, \bibinfo
  {author} {\bibfnamefont {C.~Y.}\ \bibnamefont {Zhu}},\ and\ \bibinfo {author}
  {\bibfnamefont {S.~H.}\ \bibnamefont {Lin}},\ }\bibfield  {title} {\bibinfo
  {title} {Symmetry forbidden vibronic spectra and internal conversion in
  benzene},\ }\href {https://doi.org/10.1039/C0CP00120A} {\bibfield  {journal}
  {\bibinfo  {journal} {Phys.~Chem.~Chem.~Phys.~}\ }\textbf {\bibinfo {volume}
  {12}},\ \bibinfo {pages} {14967} (\bibinfo {year} {2010})}\BibitemShut
  {NoStop}%
\bibitem [{\citenamefont {Cocchi}\ \emph
  {et~al.}(2014{\natexlab{b}})\citenamefont {Cocchi}, \citenamefont {Prezzi},
  \citenamefont {Ruini}, \citenamefont {Molinari},\ and\ \citenamefont
  {Rozzi}}]{cocchi2014}%
  \BibitemOpen
  \bibfield  {author} {\bibinfo {author} {\bibfnamefont {C.}~\bibnamefont
  {Cocchi}}, \bibinfo {author} {\bibfnamefont {D.}~\bibnamefont {Prezzi}},
  \bibinfo {author} {\bibfnamefont {A.}~\bibnamefont {Ruini}}, \bibinfo
  {author} {\bibfnamefont {E.}~\bibnamefont {Molinari}},\ and\ \bibinfo
  {author} {\bibfnamefont {C.~A.}\ \bibnamefont {Rozzi}},\ }\bibfield  {title}
  {\bibinfo {title} {Ab initio simulation of optical limiting: the case of
  metal-free phthalocyanine},\ }\href
  {https://doi.org/10.1103/PhysRevLett.112.198303} {\bibfield  {journal}
  {\bibinfo  {journal} {Phys.~Rev.~Lett.~}\ }\textbf {\bibinfo {volume}
  {112}},\ \bibinfo {pages} {198303} (\bibinfo {year}
  {2014}{\natexlab{b}})}\BibitemShut {NoStop}%
\bibitem [{\citenamefont {Guandalini}\ \emph {et~al.}(2021)\citenamefont
  {Guandalini}, \citenamefont {Cocchi}, \citenamefont {Pittalis}, \citenamefont
  {Ruini},\ and\ \citenamefont {Rozzi}}]{guandalini2021}%
  \BibitemOpen
  \bibfield  {author} {\bibinfo {author} {\bibfnamefont {A.}~\bibnamefont
  {Guandalini}}, \bibinfo {author} {\bibfnamefont {C.}~\bibnamefont {Cocchi}},
  \bibinfo {author} {\bibfnamefont {S.}~\bibnamefont {Pittalis}}, \bibinfo
  {author} {\bibfnamefont {A.}~\bibnamefont {Ruini}},\ and\ \bibinfo {author}
  {\bibfnamefont {C.~A.}\ \bibnamefont {Rozzi}},\ }\bibfield  {title} {\bibinfo
  {title} {Nonlinear light absorption in many-electron systems excited by an
  instantaneous electric field: a non-perturbative approach},\ }\href@noop {}
  {\bibfield  {journal} {\bibinfo  {journal} {Phys.~Chem.~Chem.~Phys.~}\
  }\textbf {\bibinfo {volume} {23}},\ \bibinfo {pages} {10059} (\bibinfo {year}
  {2021})}\BibitemShut {NoStop}%
\bibitem [{\citenamefont {De~Sio}\ \emph {et~al.}(2019)\citenamefont {De~Sio},
  \citenamefont {Nguyen},\ and\ \citenamefont {Lienau}}]{desi+19zna}%
  \BibitemOpen
  \bibfield  {author} {\bibinfo {author} {\bibfnamefont {A.}~\bibnamefont
  {De~Sio}}, \bibinfo {author} {\bibfnamefont {X.~T.}\ \bibnamefont {Nguyen}},\
  and\ \bibinfo {author} {\bibfnamefont {C.}~\bibnamefont {Lienau}},\
  }\bibfield  {title} {\bibinfo {title} {Signatures of strong vibronic coupling
  mediating coherent charge transfer in two-dimensional electronic
  spectroscopy},\ }\href@noop {} {\bibfield  {journal} {\bibinfo  {journal}
  {Z.~Naturforsch.~A}\ }\textbf {\bibinfo {volume} {74}},\ \bibinfo {pages}
  {721} (\bibinfo {year} {2019})}\BibitemShut {NoStop}%
\bibitem [{\citenamefont {Chen}\ \emph {et~al.}(2020)\citenamefont {Chen},
  \citenamefont {Zuehlsdorff}, \citenamefont {Morawietz}, \citenamefont
  {Isborn},\ and\ \citenamefont {Markland}}]{chen+2020jpcl}%
  \BibitemOpen
  \bibfield  {author} {\bibinfo {author} {\bibfnamefont {M.~S.}\ \bibnamefont
  {Chen}}, \bibinfo {author} {\bibfnamefont {T.~J.}\ \bibnamefont
  {Zuehlsdorff}}, \bibinfo {author} {\bibfnamefont {T.}~\bibnamefont
  {Morawietz}}, \bibinfo {author} {\bibfnamefont {C.~M.}\ \bibnamefont
  {Isborn}},\ and\ \bibinfo {author} {\bibfnamefont {T.~E.}\ \bibnamefont
  {Markland}},\ }\bibfield  {title} {\bibinfo {title} {Exploiting machine
  learning to efficiently predict multidimensional optical spectra in complex
  environments},\ }\href {https://doi.org/10.1021/acs.jpclett.0c02168}
  {\bibfield  {journal} {\bibinfo  {journal} {J.~Phys.~Chem.~Lett.}\ }\textbf
  {\bibinfo {volume} {11}},\ \bibinfo {pages} {7559} (\bibinfo {year}
  {2020})},\ \bibinfo {note} {pMID: 32808797},\ \Eprint
  {https://arxiv.org/abs/https://doi.org/10.1021/acs.jpclett.0c02168}
  {https://doi.org/10.1021/acs.jpclett.0c02168} \BibitemShut {NoStop}%
\bibitem [{\citenamefont {Xue}\ \emph {et~al.}(2020)\citenamefont {Xue},
  \citenamefont {Barbatti},\ and\ \citenamefont {Dral}}]{xue+2020jpca}%
  \BibitemOpen
  \bibfield  {author} {\bibinfo {author} {\bibfnamefont {B.-X.}\ \bibnamefont
  {Xue}}, \bibinfo {author} {\bibfnamefont {M.}~\bibnamefont {Barbatti}},\ and\
  \bibinfo {author} {\bibfnamefont {P.~O.}\ \bibnamefont {Dral}},\ }\bibfield
  {title} {\bibinfo {title} {Machine learning for absorption cross sections},\
  }\href@noop {} {\bibfield  {journal} {\bibinfo  {journal} {J.~Phys.~Chem.~A}\
  }\textbf {\bibinfo {volume} {124}},\ \bibinfo {pages} {7199} (\bibinfo {year}
  {2020})}\BibitemShut {NoStop}%
\bibitem [{\citenamefont {Min}\ \emph {et~al.}(2015)\citenamefont {Min},
  \citenamefont {Agostini},\ and\ \citenamefont {Gross}}]{min+2015prl}%
  \BibitemOpen
  \bibfield  {author} {\bibinfo {author} {\bibfnamefont {S.~K.}\ \bibnamefont
  {Min}}, \bibinfo {author} {\bibfnamefont {F.}~\bibnamefont {Agostini}},\ and\
  \bibinfo {author} {\bibfnamefont {E.~K.~U.}\ \bibnamefont {Gross}},\
  }\bibfield  {title} {\bibinfo {title} {Coupled-trajectory quantum-classical
  approach to electronic decoherence in nonadiabatic processes},\ }\href
  {https://doi.org/10.1103/PhysRevLett.115.073001} {\bibfield  {journal}
  {\bibinfo  {journal} {Phys.~Rev.~Lett.~}\ }\textbf {\bibinfo {volume}
  {115}},\ \bibinfo {pages} {073001} (\bibinfo {year} {2015})}\BibitemShut
  {NoStop}%
\bibitem [{\citenamefont {Curchod}\ \emph {et~al.}(2018)\citenamefont
  {Curchod}, \citenamefont {Agostini},\ and\ \citenamefont
  {Tavernelli}}]{curc+2018epjb}%
  \BibitemOpen
  \bibfield  {author} {\bibinfo {author} {\bibfnamefont {B.~F.~E.}\
  \bibnamefont {Curchod}}, \bibinfo {author} {\bibfnamefont {F.}~\bibnamefont
  {Agostini}},\ and\ \bibinfo {author} {\bibfnamefont {I.}~\bibnamefont
  {Tavernelli}},\ }\bibfield  {title} {\bibinfo {title} {Ct-mqc -- a
  coupled-trajectory mixed quantum/classical method including nonadiabatic
  quantum coherence effects},\ }\href
  {https://doi.org/10.1140/epjb/e2018-90149-x} {\bibfield  {journal} {\bibinfo
  {journal} {Eur.~Phys.~J.~B}\ }\textbf {\bibinfo {volume} {91}},\ \bibinfo
  {pages} {168} (\bibinfo {year} {2018})}\BibitemShut {NoStop}%
\bibitem [{\citenamefont {Gossel}\ \emph {et~al.}(2018)\citenamefont {Gossel},
  \citenamefont {Agostini},\ and\ \citenamefont {Maitra}}]{gossel+2018jctc}%
  \BibitemOpen
  \bibfield  {author} {\bibinfo {author} {\bibfnamefont {G.~H.}\ \bibnamefont
  {Gossel}}, \bibinfo {author} {\bibfnamefont {F.}~\bibnamefont {Agostini}},\
  and\ \bibinfo {author} {\bibfnamefont {N.~T.}\ \bibnamefont {Maitra}},\
  }\bibfield  {title} {\bibinfo {title} {Coupled-trajectory mixed
  quantum-classical algorithm: A deconstruction},\ }\href
  {https://doi.org/10.1021/acs.jctc.8b00449} {\bibfield  {journal} {\bibinfo
  {journal} {J.~Chem.~Theory.~Comput.~}\ }\textbf {\bibinfo {volume} {14}},\
  \bibinfo {pages} {4513} (\bibinfo {year} {2018})},\ \bibinfo {note} {pMID:
  30063343},\ \Eprint
  {https://arxiv.org/abs/https://doi.org/10.1021/acs.jctc.8b00449}
  {https://doi.org/10.1021/acs.jctc.8b00449} \BibitemShut {NoStop}%
\end{thebibliography}

%

\end{document}